\documentclass[a4paper,12pt]{article}
 
\pdfoutput=1
\usepackage{jheppub}
\usepackage{dcolumn}
\usepackage{soul}
\usepackage[english]{babel}
\usepackage[utf8]{inputenc}
\usepackage{dcolumn}
\usepackage{soul}
\usepackage{hyperref}
\usepackage{url}
\usepackage{graphicx}
\usepackage{bm,amsmath,amssymb}
\usepackage[mathscr]{eucal}
\usepackage{makeidx}
\usepackage{subfig}
\usepackage{amsmath}
\usepackage{amssymb}
\usepackage{amsthm}
\usepackage{mathrsfs}
\usepackage{graphicx}
\usepackage{fancyhdr}
\usepackage{array}
\usepackage{simplewick}
\usepackage{latexsym}
\usepackage[all]{xy}
\usepackage{enumerate}
\usepackage{dsfont}
\usepackage{slashed}
\usepackage{float}
\usepackage[titletoc]{appendix}
\usepackage{ulem}

\usepackage{verbatim}

\usepackage{tikz,tikz-3dplot}
\tdplotsetmaincoords{80}{45}

\tikzset{surface/.style={draw=black, fill=white, fill opacity=.6}}

\usepackage{tikz}
\usetikzlibrary{decorations.pathmorphing,patterns}

\newcommand{\be}{\begin{equation}}
\newcommand{\ee}{\end{equation}}
\newcommand{\bea}{\begin{eqnarray}}
\newcommand{\eea}{\end{eqnarray}}

\newcommand{\ben}{\begin{eqnarray}}
\newcommand{\een}{\end{eqnarray}}

\title{ Holographic Boundary Conformal Field Theory   within 
Horndeski Gravity}

\author{Fabiano F. Santos$^{a}$, Behnam Pourhassan$^{b,c}$, Emmanuel N. 
Saridakis$^{f,g,h}$, Oleksii Sokoliuk$^{i,j}$, Alexander Baransky$^{j}$, and 
Emre Onur Kahya$^{d}$}
\affiliation{$^{a}$Instituto de F\'{\i}sica, Universidade Federal do Rio de 
Janeiro, 21.941-909, Rio de Janeiro, RJ, Brazil.\\Departamento de Física, 
Universidade Federal do Maranhão, São Luís, 65080-805, Brazil.}
\affiliation{$^{b}$School of Physics, Damghan University, Damghan, 3671641167, 
Iran.}
\affiliation{$^{c}$Center for Theoretical Physics, Khazar University, 41 
Mehseti 
Street, Baku, AZ1096, Azerbaijan.}
\affiliation{$^{d}$Physics Department, Istanbul Technical University, Istanbul 34469, Turkey.}
\affiliation{$^{f}$Institute for Astronomy,  Astrophysics, Space Applications 
and 
Remote Sensing (IAASARS), National Observatory of Athens, Athens, Greece.}

\affiliation{$^{g}$
CAS Key Laboratory for Research in Galaxies and Cosmology, University of 
Science 
and Technology of China, Hefei, Anhui 230026, China.}
\affiliation{$^{h}$Departamento de Matem\'{a}ticas, Universidad Cat\'{o}lica 
del 
Norte, Avda. Angamos 0610, Casilla 1280 Antofagasta, Chile}
\affiliation{$^{i}$ Astronomical Observatory, Taras Shevchenko National 
University of Kyiv, 3 Observatorna St., 04053 Kyiv, Ukraine,}
\affiliation{$^{j}$Main Astronomical Observatory of the NAS of Ukraine (MAO 
NASU), Kyiv, 03143, Ukraine.}

\abstract{We  investigate  entanglement islands and the Page curve in the framework of  
Horndeski gravity on a Karch-Randall braneworld background. In particular, 
treating  the holographic boundary conformal field theory analytically 
 we find that the Horndeski parameters significantly alter the 
behavior of the Page curve compared to standard general relativity, a feature 
caused by the nontrivial geometry induced by the  Horndeski scalar field.  
Interestingly enough,  the geometry far from the AdS limit plays a more 
significant role compared to previous studies. This suggests that Horndeski gravity 
introduces important modifications to the distribution of quantum information in the 
holographic model. Finally, we claim that holographic consistency can be used
reversely to impose constraints on Horndeski gravity itself, providing a   
new tool for probing the validity of modified gravity theories.  }

\begin{document}
	\maketitle
	\newcommand{\limit}[3]
	{\ensuremath{\lim_{#1 \rightarrow #2} #3}}
    
\section{Introduction}

The black hole information loss paradox has been a central issue in theoretical 
physics since its introduction, gaining particular prominence following the 
insights provided by Page curve.
This curve suggests a potential violation of unitarity when the fine-grained 
entropy of Hawking radiation from an
evaporating black hole surpasses the entropy of the black hole itself 
\cite{Hawking:1975vcx,Hawking:1976ra}.
In response, a variety of alternative theoretical models have been proposed to 
address this paradox more effectively.
Among these, the concept of holographic complexity has emerged as a significant 
approach 
\cite{Susskind:2014rva,Brown:2015bva,Lloyd:2000cry,Brown:2015lvg,
Susskind:2018fmx,Brown:2018bms,Brown:2017jil,Brown:2019whu,Brown:2022rwi}.
This framework posits that black holes continue to emit information even after 
reaching thermal equilibrium.

Recent investigations have further explored these
issues within the context of modified gravity theories, particularly Horndeski 
gravity \cite{Horndeski:1974wa, 
Kobayashi:2019hrl,Lu:2020iav,Kase:2018aps,Koyama:2013paa,Bahamonde:2021dqn,
Petronikolou:2021shp}, which is a significant subject knowing the theoretical 
and observational advantages of gravitational modifications 
\cite{CANTATA:2021asi,Abdalla:2022yfr,Capozziello:2011et,Cai:2015emx}.
Studies such as those by 
\cite{Santos:2021orr,Sokoliuk:2022llp,Santos:2023flb,Santos:2023mee}
have demonstrated that residual information can be extracted by post-thermal 
equilibrium through 
the Anti-de-Sitter/Boundary Conformal Field Theory (AdS/BCFT) correspondence.
This correspondence underscores the critical role of black hole entropy 
(S$_{BH}$) \cite{Santos:2024zoh} in 
the black hole information paradox, offering a promising avenue for further 
research into the intricate relationship
between black hole thermodynamics and holographic principle.

Hence, developments in holographic transport and gravity-related quantum 
effects 
present a comprehensive overview of the intricate connections between the 
Anti-de-Sitter/Conformal Field Theory (AdS/CFT) correspondence and modified 
gravity theories, particularly Horndeski gravity 
\cite{Maldacena:1997re,Witten:1998qj,Caceres:2023gfa,Santos:2024cwf,Takayanagi:2011zk,
Fujita:2011fp,Kanda:2023zse,dosSantos:2022scy}. 
Moreover, 
the extension of AdS/CFT proposed by Takayanagi, which emphasizes boundary 
effects and
their relation to entanglement entropy \cite{Takayanagi:2011zk, Fujita:2011fp, 
Ryu:2006bv, DosSantos:2022exb,Caceres:2017lbr,Feng:2015oea,Wald:1993nt,Iyer:1994ys}, 
opens promising avenues for exploring the interplay between quantum
gravity and field theories 
\cite{Karch:2000ct,DeWolfe:2001pq,Bak:2003jk,Clark:2004sb,
Cardy:2004hm,Tonni:2010pv,Azeyanagi:2007qj, 
Setare:2008hm,Saridakis:2017rdo,Basilakos:2023seo}.
The discussion of Hawking-Page phase transitions and corrections to boundary 
entropy underscores the richness of the holographic 
framework in addressing complex phenomena. By describing gravity duals within 
this context enlightens the discussion on the relation 
  between quantum information theory and gravitational physics, 
particularly in scenarios where traditional
quantum gravity approaches may be inadequate.

\begin{figure}[!ht]
\vspace{-1cm}
\begin{center}
\includegraphics[width=\textwidth]{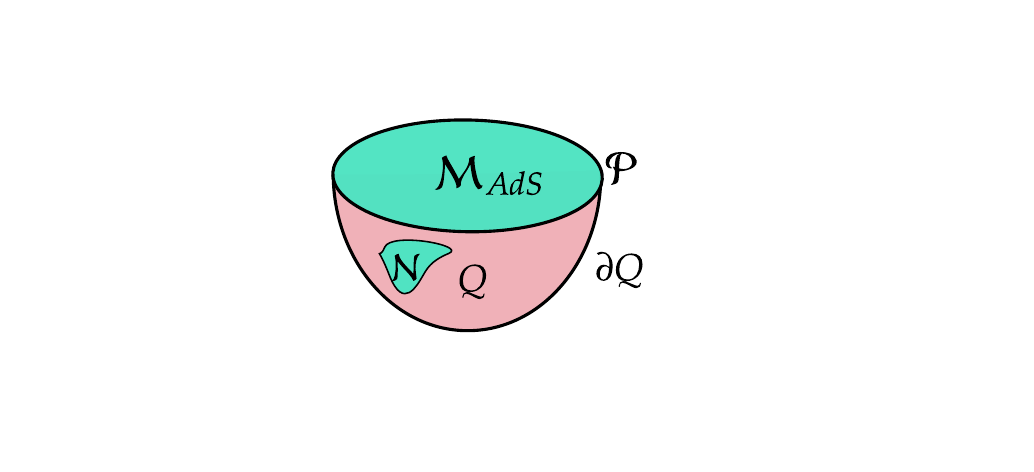}
\vspace{-3cm}
\end{center}
\caption{{\it{Shcematic representation of AdS/BCFT correspondence. Here, 
$\partial\mathcal{N} =\mathcal{M}\cup\,Q$, where $Q$ is a $d$-dimensional 
manifold   satisfying $ \partial Q\cap\partial \mathcal{M}=\mathcal{P}$. 
}}}\label{P}
\end{figure}

We mention here that the entanglement entropy in the context of Horndeski 
gravity 
\cite{Ryu:2006bv,DosSantos:2022exb}
aligns with the boundary entropy in two-dimensional BCFTs 
\cite{Santos:2021orr}. 
In these frameworks, 
gravity duals for Einstein and Horndeski gravity are defined at the CFT 
boundary 
on the AdS$_d$-dimensional 
manifold $\mathcal{M}$, which is asymptotically 
AdS$_{d+1}$-dimensional.
For visualization we 
refer to Fig. \ref{P} to elucidate the topological aspects of the AdS space, 
where as we can see  the involved boundaries provides a concrete foundation for 
the discussion, 
enhancing the understanding of these complex theoretical constructs.

The recent discussions on doubly holographic descriptions and the island 
formula reveal the interplay between holography, entanglement, and gravity.
The use of BCFTs to understand these 
concepts 
is a significant advancement, especially in the context of understanding how 
entanglement entropy behaves in gravitational settings 
\cite{Almheiri:2019hni,Rozali:2019day,Chen:2020uac,Chen:2020hmv,Deng:2020ent,
Suzuki:2022xwv,Geng:2021iyq,Geng:2024xpj,Geng:2020fxl}. 
The framework we are outlining, where a CFT coupled to semiclassical gravity 
provides a dual description of a (d+1)-dimensional theory, 
is a robust platform for exploring entanglement phenomena. Additionally, the 
Ryu-Takayanagi 
(RT) prescription is central to this exploration \cite{Ryu:2006ef,Ryu:2006bv}, 
enabling the calculation of holographic entanglement entropy in terms of 
geometric quantities in the gravitational dual \cite{Suzuki:2022xwv}.

\begin{figure}[!ht]
\begin{center}
\includegraphics[width=\textwidth]{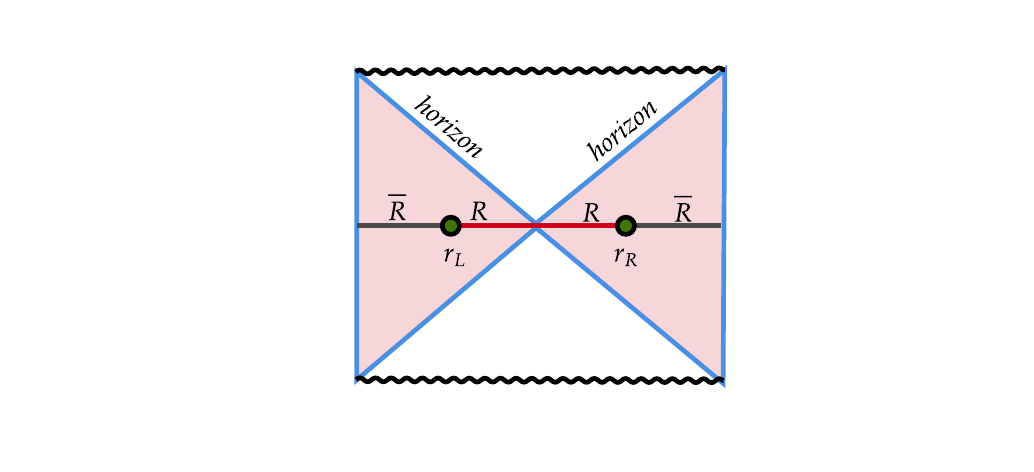}
\vspace{-2cm}
\end{center}
\caption{{\it{Subsystems Penrose diagram (see text).}}}\label{p1}
\end{figure}

Our approach, deriving the island formula within Horndeski gravity and 
connecting it to work on holographic entanglement entropy in disjoint 
subsystems, offers insight to these entanglement measures. Hence, the 
treatment of subsystems $A$ (subregion $\mathcal{R}$) and its complement 
$\bar{\mathcal{R}}$,  and their connection through the bifurcation surface, 
illustrates how the RT formula can be adapted in more complex gravitational 
settings \cite{Basu:2023jtf}. The visualization using the Penrose diagram (see 
Fig. 
\ref{p1}) to illustrate the relation between past and future horizons at the 
bifurcation point emphasizes the geometric intuition behind the island formula 
and its relevance in quantum gravity scenarios. Furthermore, this approach 
helps 
to clarify the 
island formula's implications and underlines the importance of horizon dynamics 
in the context of entanglement and information loss. In particular, in this 
scenario we 
calculate the entropy of the subregion $\mathcal{R}$, which is connected 
internally through the bifurcation where the horizon at $t=0$ separates the 
past 
and future horizons (this is the point in the center of the Penrose diagram in 
Fig. 
\ref{p1}).

The exploration of gravitational duals in AdS$_{d+1}$ through the framework of 
Karch-Randall (KR) branes presents a compelling approach to understanding the 
interplay between gravity on the brane and non-gravitational baths via 
holography 
\cite{Randall:1999vf,Karch:2000ct,Brito:2018pwe,Santos:2023eqp,Almheiri:2019psf,
Penington:2019npb,Engelhardt:2014gca,Almheiri:2019hni,Almheiri:2019psy}. This 
connection to AdS$_{d}$ gravity, where the brane is coupled to a  CFT, serves 
as a powerful tool for probing the dynamics of the system, 
particularly  the entanglement entropy and information flow. Thus, the 
new input gained from Horndeski gravity, particularly the reduction of brane 
physics to entanglement entropy, is useful to understand the  
local 
physics 
on the brane \cite{Karch:2000ct}. Traditional limits, such as those in disk 
AdS$_{4}$-dS$_{4}$, may not fully encapsulate the subtleties necessary to 
understand local phenomena comprehensively.

In the context of 
\cite{Santos:2023eqp}, dividing mass excitations into two distinct sets 
provides 
a new perspective on the distribution and processing of quantum information 
between the CFT on the brane and the bulk dynamics. This separation has 
significant implications for understanding how information is stored and 
retrieved in these holographic models, especially in scenarios involving black 
holes and entropy dynamics.

The analysis of the interplay between a CFT$_d$ coupled to AdS$_d$ gravity on 
the brane, particularly in the context of transparent boundary conditions, 
offers  information on the dynamics of such systems 
\cite{Basu:2023jtf,Hartman:2013qma, Geng:2024xpj,Geng:2021mic,Geng:2023qwm,Geng:2021hlu,
Geng:2023zhq,Geng:2022dua}. These boundary conditions serve as a crucial 
interface between the brane dynamics and the non-gravitational bath, leading to 
significant physical phenomena, such as the effective mass acquisition of the 
graviton due to energy exchange 
\cite{Karch:2000ct,Brito:2018pwe,Santos:2023eqp}. The role of transparent 
boundary conditions in facilitating this mass acquisition is important, as 
highlighted in works like \cite{Geng:2021mic}. This mechanism illustrates how 
the gravitational theory on the brane can influence bulk dynamics, thereby 
affecting the properties of gravitational excitations. 

A significant observation is that these boundary conditions result in the 
non-conservation of 
the stress tensor within the BCFT defect \cite{Santos:2023mee}. This 
non-conservation can be interpreted as a reflection of the dynamics of quantum 
information and entanglement in these settings, which is essential for 
understanding the implications of residual information within the Horndeski 
framework. Furthermore, linking boundary entropy to this residual information 
and its role in the growth of entanglement entropy reveals the   
relations  present in these holographic scenarios 
\cite{Santos:2021orr,Santos:2023mee}. Thus, investigating how these two classes 
of 
surfaces contribute to the growth of entanglement entropy  offers information 
on the relation between boundary conditions, entanglement dynamics, and 
gravitational 
theories.
 
Recently, some works investigated   the entanglement islands and the Page curve 
within the framework of Horndeski 
\cite{Geng:2021mic,Geng:2023qwm,Geng:2021hlu}.   Our goal  to 
explore entanglement islands 
within a Horndeski gravitational framework, focusing on a single scalar field 
that induces symmetry breaking,  
aligns with studies suggesting that weakly broken symmetries can violate the 
area theorem, regardless of whether the symmetry breaking is explicit or 
spontaneous \cite{Bass:2021acr,Higgs2014,Jeong:2022zea}.   As indicated in 
\cite{Santos:2021orr,Sokoliuk:2022llp,Santos:2023flb,Santos:2023mee}, the 
interplay between symmetry breaking and residual entropy provides crucial 
insights into the nature of quantum states in gravitational contexts. Hence, 
these studies are useful concerning  quantum gravity and 
information theory, particularly within frameworks that account for symmetry 
breaking.

This work is summarized as follows. In Sec. \ref{v0} we present the doubly 
holographic black string and we extract the conditions that this solution 
satisfies to become a solution of the Horndeski gravity. In Sec. \ref{ETER}  we 
calculate the density functional to provide the entanglement thermodynamics in 
Horndeski gravity, and   we extract the entanglement entropy through 
spatial embedding formalism and Field Theory computation, while in Sec. 
\ref{T-A-PAGE} we present the Page curve behavior. Finally, Sec. \ref{CONCL} 
provides our conclusions.  

\section{Horndeski gravity on doubly holographic black string}\label{v0}

In order to investigate the structure of the AdS$_{4}$/BCFT$_{3}$ 
correspondence, we start by embedding a three-dimensional Karch-Randall 
(KR) brane  \cite{Randall:1999vf,Karch:2000ct} 
in a four-dimensional black string. The proposal to 
analyze this 
scenario within the context of  Horndeski gravity allows for an 
investigation of the
effects of modified gravity on the system  behavior, particularly regarding 
thermal states 
and the degrees of freedom associated with the black hole background.
Hence, we consider the following  ansatz of a four-dimensional black 
string, 
truncated by a Karch-Randall brane, described by the metric
\begin{equation}
ds^{2}_{AdS_4}=\frac{1}{r^{2}\sin^{2}(u)}\left(-f(r)dt^{2}+dy^{2}+r^{2}du^{2}
+\frac{dr^{2}}{f(r)}\right).\label{me}
\end{equation}
In our prescription $u\,\epsilon\,[-\infty,\infty]$ and
$u\,=-\infty\,\cup\,\infty$ is the asymptotic boundary and 
the $KR$ brane is embedded at a constant $u=u_b$ slice \cite{Basu:2023jtf}.

\begin{figure}[!ht]
\begin{center}
\includegraphics[width=\textwidth]{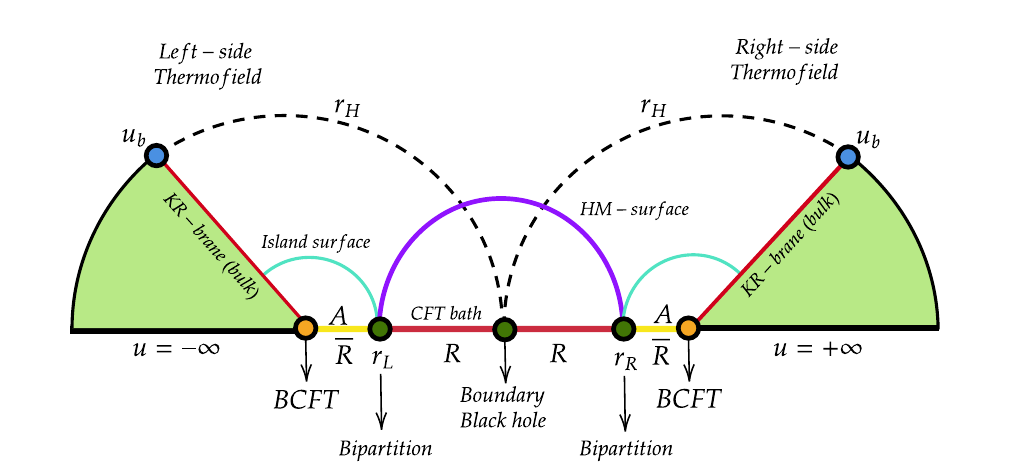}
\end{center}
\caption{{\it{Schematic representation of the black string. The $KR$ brane is 
the point $u=u_b$ (red line), the dashed arc at the $r=r_{H}$ represents the black 
string horizon, and the violet arc is a Hatman-Maldacena (HM) surface 
\cite{Hartman:2013qma}. Finally, the island surfaces correspond to the blue 
curves.}}}\label{p2}
\end{figure}

In Fig. \ref{p2} we present a schematic representation of the black string.
The accessible bulk region extends from $u=u_b$ to $u=\infty$. As we   
observe, the geometry in each constant-$u$ slice of the black string is an eternal AdS$_4$ black hole 
with asymptotic limits. From the overview point of AdS$_4$/BCFT$_3$ correspondence, the dual field 
theory is a BCFT$_3$ in an AdS$_3$ black hole background with conformal 
boundary conditions at $r=0$ with the following metric
\begin{eqnarray}
ds^{2}_{AdS_3}=\frac{1}{r^{2}}\left(-f(r)dt^{2}+dy^{2}+\frac{dr^{2}}{f(r)}
\right)\,.\label{Q1} 
\end{eqnarray}

Our scenario incorporates Karch-Randall branes $Q^{L}\,\cup\,Q^{R}$ 
(i.e.  with asymptotically AdS$_3$ geometries) into an asymptotically AdS$_3$
volume described by the following action 
\begin{eqnarray}
&&
\!\!\!\!\!\!
S=\kappa\int_{\mathcal{N}}{d^{4}x\sqrt{-g}\mathcal{L}_{H}}+S^{\mathcal{N}}_{
mat}+2\kappa\int_{Q^{L}\,\cup\,Q^{R}}{d^{3}x\sqrt{-h}\mathcal{L}_{bdry}}\cr
&& \ \  \,
+2\int_{Q^{L}\,\cup\,Q^{R}}{d^{3}x\sqrt{-h}\mathcal{L}_{mat}}+2\kappa\int_{Q^{
L}\,\cup\,Q^{R},ct}{d^{3}x\sqrt{-h}\mathcal{L}_{ct}}\,, \label{S}
\end{eqnarray}
where
  $\mathcal{N}$ are the two branes form the boundary of a wedge (see Fig. 
\ref{p2}) 
\cite{Basu:2023jtf,Geng:2021mic,Geng:2023qwm,Geng:2021hlu,Geng:2023zhq,
Geng:2022dua,Saridakis:2007ns,Kofinas:2014qxa}. 
Furthermore,  $S^{\mathcal{N}}_{mat}$ describes 
ordinary matter that 
is supposed to be a perfect fluid, and
$\mathcal{L}_{mat}$ is a Lagrangian of possible matter fields on 
$Q^{L}\,\cup\,Q^{R}$.
In the above action we have 
introduced the specific but quite general Horndeski Lagrangians 
\cite{Santos:2021orr,Santos:2023flb,Santos:2023mee}:
\begin{eqnarray}
&&{\cal L}_H=(R-2\Lambda)-\frac{1}{2}(\alpha g_{\mu\nu}-\gamma\,  
G_{\mu\nu})\nabla^{\mu}\phi\nabla^{\nu}\phi\,,\\
&&\mathcal{L}_{bdry}=(K-\Sigma)+\frac{\gamma}{4}(\nabla_{\mu}\phi\nabla_{\nu}
\phi 
n^{\mu}n^{\nu}-(\nabla\phi)^{2})K+\frac{\gamma}{4}\nabla_{\mu}\phi\nabla_{\nu}
\phi K^{\mu\nu}\,, \label{3} 
\end{eqnarray}
where $\phi$ is the Horndeski scalar field and   
$\alpha$ and $\gamma$ are the     Horndeski parameters.
In the boundary Lagrangian \eqref{3}, 
$K_{\mu\nu}=h^{\beta}_{\mu}\nabla_{\beta}n_{\nu}$ is the extrinsic curvature, 
$h_{\mu\nu}$ is the induced metric and $n^\mu$ is the normal vector of the 
hypersurface $Q^{L}$($Q^{R}$). The traceless contraction of $K_{\mu\nu}$ is 
$K=h^{\mu\nu}K_{\mu\nu}$, and $\Sigma^{L}$($\Sigma^{R}$) is the boundary 
tension 
on $Q^{L}$($Q^{R}$). Finally, 
${\cal L}_{ct}$ are boundary counterterms 
localized on $\mathcal{P}$ \cite{Santos:2021orr,Santos:2023flb,Santos:2023mee}, 
which is required to be an asymptotic AdS 
spacetime, and are given by
\begin{eqnarray} 
&&\mathcal{L}_{ct}=c_{0}+c_{1}R+c_{2}R^{ij}R_{ij}+c_{3}R^{2}+b_{1}(\partial_{i}
\phi\partial^{i}\phi)^{2}+....\label{4}
\end{eqnarray}

 Assuming that $S^{\mathcal{N}}_{mat}$ is constant, and varying $S^\mathcal{N}$ 
with respect to $g_{\alpha\beta}$ and $\phi$, we obtain the field equations 
\begin{eqnarray}
{\cal E}_{\mu\nu}[g_{\mu\nu},\phi]&=&G_{\mu\nu}+\Lambda 
g_{\mu\nu}-\frac{\alpha}{2}\left(\nabla_{\mu}\phi\nabla_{\nu}\phi-\frac{1}{2}g_{
\mu\nu}\nabla_{\lambda}\phi\nabla^{\lambda}\phi\right)\label{11}\nonumber\\    
           &-&\frac{\gamma}{2}\left(\frac{1}{2}\nabla_{\mu}\phi\nabla_{\nu}\phi 
R-2\nabla_{\lambda}\phi\nabla_{(\mu}\phi 
R^{\lambda}_{\nu)}-\nabla^{\lambda}\phi\nabla^{\rho}\phi 
R_{\mu\lambda\nu\rho}\right)\nonumber\\									
&-&\frac{\gamma}{2}\left(-(\nabla_{\mu}\nabla^{\lambda}\phi)(\nabla_{\nu}\nabla_
{\lambda}\phi)+(\nabla_{\mu}\nabla_{\nu}\phi)\Box\phi+\frac{1}{2}G_{\mu\nu}
(\nabla\phi)^{2}\right)\nonumber\\
									&+&\frac{\gamma 
g_{\mu\nu}}{2}\left(-\frac{1}{2}(\nabla^{\lambda}\nabla^{\rho}\phi)(\nabla_{
\lambda}\nabla_{\rho}\phi)+\frac{1}{2}(\Box\phi)^{2}-(\nabla_{\lambda}
\phi\nabla_{\rho}\phi)R^{\lambda\rho}\right),\\
{\cal E}_{\phi}[g_{\mu\nu},\phi]&=&\nabla_{\mu}\left[\left(\alpha 
g^{\mu\nu}-\gamma G^{\mu\nu}\right)\nabla_{\nu}\phi\right]\,.\label{12}
\end{eqnarray}
The no-hair theorem requires that the square of the radial component of  
the conserved current vanishes identically without restricting the radial 
dependence
of the scalar field, which implies 
\begin{equation}
\alpha g_{rr}-\gamma G_{rr}=0\label{NO-HAIR}\,,
\end{equation}
and from this condition we find    ${\cal E}_{\phi}[g_{rr},\phi]=0$ 
\cite{Santos:2021orr,Santos:2023flb,Santos:2023mee}. Without loss of 
generality we consider   
$\phi=\phi(r)$ and we define  $\phi^{'}(r)\equiv\psi(r)$. 
As it can be shown, the equations ${\cal E}_{\phi}[g_{rr},\phi]={\cal 
E}_{rr}[g_{rr},\phi]=0$ are satisfied, and 
therefore we can calculate the horizon functions 
$f(r)$ and $\psi(r)$  as (for the AdS$_4$ radius we set 
1) 
\begin{eqnarray}
f(r)&=&1+Cr^{3}+\left(-3+\frac{\alpha}{\gamma}\right)\csc^2(u),\label{P1}\\
\psi^{2}(r)&=&-\frac{4(\alpha+\gamma\Lambda)}{\alpha\gamma 
r^{2}A(u)\sin^{2}(u)}\frac{1}{f(r)}.\label{P2}\\
A(u)&=&2\alpha-5\gamma-\gamma\cos(2u),
\end{eqnarray}
where   equations (\ref{P1})-(\ref{P2}) are found though   
(\ref{NO-HAIR}) with $r>0$, $0<u<\pi$  and $x,y\in \mathbb{R}$. The blackening 
factor $f(r)$ gives a black hole on each constant-$u$ slice, and one such slice 
will be the  KR brane \cite{Randall:1999vf,Karch:2000ct}. 
Finally, replacing (\ref{P1}) and (\ref{P2}) in   Horndeski equations of motion, 
we find the following conditions
\begin{eqnarray}
&&\frac{(\beta_0+1)}{A(u)}=0\label{Li}\\
&&\frac{(\beta_0+1)}{(\alpha-\gamma)}-\frac{(\beta_0+1)}{A(u)}=0\label{Lii},
\end{eqnarray}
with $\beta_0=\alpha/\gamma\Lambda$, and where the parameters are defined in 
the range $-\infty<\beta_0\leq-1$  with $\alpha,\gamma<0$, or $-1\leq\beta_0<0$  with 
$\alpha,\gamma>0$. These conditions provide $\beta_0+1\geq\,0$, which shows a 
black string solution for Horndeski gravity where the scalar field does not 
vanish \cite{Li:2018kqp}. Finally, the two branes are located at $y(r)$ 
hypersurface, described by
\begin{eqnarray}
&&\left[K_{\alpha\beta}-h_{\alpha\beta}(K-\Sigma)+\frac{\gamma}{4}H_{\alpha\beta
}\right]_{Q^{L}\,\cup\,Q^{R}}=0\label{5}\\
&&H_{\alpha\beta}\equiv(\nabla_{\mu}\phi\nabla_{\nu}\phi\,n^{\mu}n^{\nu}
-(\nabla\phi)^{2})(K_{\alpha\beta}-h_{\alpha\beta}K)-(\nabla_{\mu}\phi\nabla^{
\mu}\phi)h_{\alpha\beta}K\,,\label{6}
\end{eqnarray}
where the defect is given 
\begin{eqnarray}
y{'}(r)&=&\frac{(\Sigma)}{\sqrt{1-\dfrac{\xi}{1-\left(\frac{r}{r_{h}}\right)^{2}
}-(\Sigma)^{2}\left(1-\left(\frac{r}{r_{h}}\right)^{2}\right)}}\,;\,\xi=-\frac{1
}{2}\frac{\alpha+\gamma\Lambda}{\alpha}.
\label{19}
\end{eqnarray}
We mention here that  for Horndeski gravity  we have a restriction on the 
parameters of the theory due to the lack of a consistent solution within this 
theory. On the other hand, for the case of Dvali-Gabadadze-Porrati (DGP) 
scenario \cite{Dvali:2000hr}  the entanglement island is not affected 
by the deformation of its configuration, thus providing constraints on the DGP 
 model through holographic consistency.  \cite{Heisenberg:2018vsk,Geng:2023qwm}.

The analysis of the island region surrounding the black hole horizon within the 
framework of Horndeski gravity can be useful for the study 
of quantum information dynamics and entanglement \cite{Geng:2021mic}. The 
observation that the defect cannot be accessed from the island without 
traversing its complement reveals the topological and geometrical 
complexities inherent in these scenarios, as illustrated in Fig. \ref{p2}. 
Finally, note that the 
 size of the island is contingent upon the choice of brane 
angle  $u_b$.

\begin{figure}[ht]
\begin{center}
\vspace{0.4cm}
\includegraphics[scale=0.45]{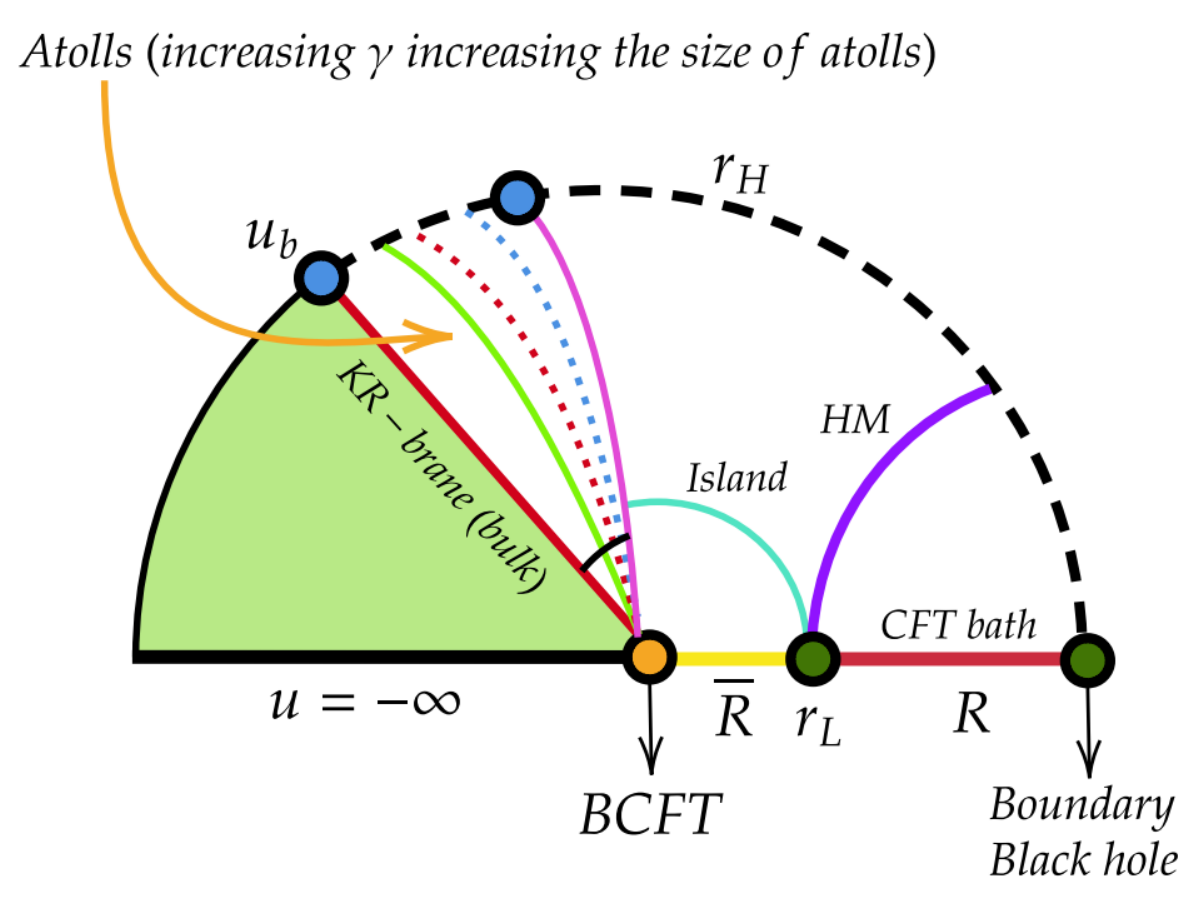}
\caption{{\it{The  relevant  Ryu-Takayanagi (RT) surfaces with $\alpha=-8/3$ 
and $\Lambda=-3$ for different values of the $\gamma$-Horndeski parameter, namely
$\gamma=-0.1$ ({\sl solid}), $\gamma=-0.2$ ({\sl dashed}), $\gamma=-0.3$ ({\sl 
dot dashed}), and $\gamma=-0.4$ ({\sl thick}). The islands are always attached 
to the brane inside the atoll; they start at the point $r_L$ and extend to the 
horizon of the black hole. Decreasing $\gamma$ reduces the size of the atoll, 
which shrinks towards the horizon, while increasing the size of $R$, limited by 
$r_L$, pushes the anchors towards the black hole  horizon. 
Finally, increasing the angle $u_b$ of the brane while keeping the Horndeski 
parameters fixed can push the anchors toward the defect.}}}
\label{ATOLL}
\end{center}
\vspace{-0.2cm}
\end{figure}

Let us now examine the different classes of Ryu-Takayanagi (RT)  surfaces 
\cite{Geng:2021mic} in Horndeski gravity. The formation of the largest 
possible island, when the island surface originates at an anchor point $r_{L}$ 
and extends to the defect (see Fig. \ref{ATOLL}), reveals  the  role 
of boundary conditions and spatial geometry in affecting the entanglement 
structure. When $r_{L}$ coincides with the defect, resulting in the atoll 
enveloping the brane entirely, it marks a significant scenario where 
entanglement properties are maximized. This suggests a profound connection 
between the island geometry, the defect location, and the resultant 
entanglement entropy.

At the boundary $r=0$ with $\Sigma=\cos(u_b)$,   Eq. (\ref{19})  becomes 
\begin{eqnarray}
y(r)=u_{0}+\frac{r\cos(u_b)}{\sqrt{-\xi}},\label{19.1}
\end{eqnarray}
and considering  for instance $u_b=\pi/2$ or $3\pi/2$, we acquire
$y(r)=u_{0}={\rm constant}$. The intersection of the branes at the boundary $r=0$ marks their 
critical role in defining the asymptotic structure of spacetime. The presence of 
positive tension on both branes, denoted as $\Sigma^{L}$($\Sigma^{R}$), indicates their 
influence on the geometry, effectively constraining the central bulk region 
between them and determining the overall gravitational configuration. As the 
branes approach their corresponding limits, the left brane at $y=0$ and the right 
brane at $y=\pi$ establish natural boundary conditions for the dual field theory. The 
positive tension suggests that the branes exert a repulsive force against the 
bulk, potentially amplifying curvature effects in their vicinity and impacting 
the properties of possible quantum fields residing on the branes.

 The exploration of the implications of Dirichlet and 
Neumann boundary conditions is important for understanding the
construction of RT surfaces, and can lead to different physical interpretations regarding the flow of 
information and the stability of entangled states in the boundary theory. 
Nevertheless, the   different angles of the brane, particularly 
below the critical angle (see Fig. \ref{ATOLL}), introduces  more complications.
In particular, the critical angle can mark a transition 
in the behavior of the RT surfaces, possibly affecting their stability and the 
nature of the entanglement islands, and surfaces that connect to the brane at 
angles below this critical threshold might exhibit different entanglement 
properties than those that do not.

\begin{figure}[!ht]
\begin{center}
\includegraphics[scale=0.95]{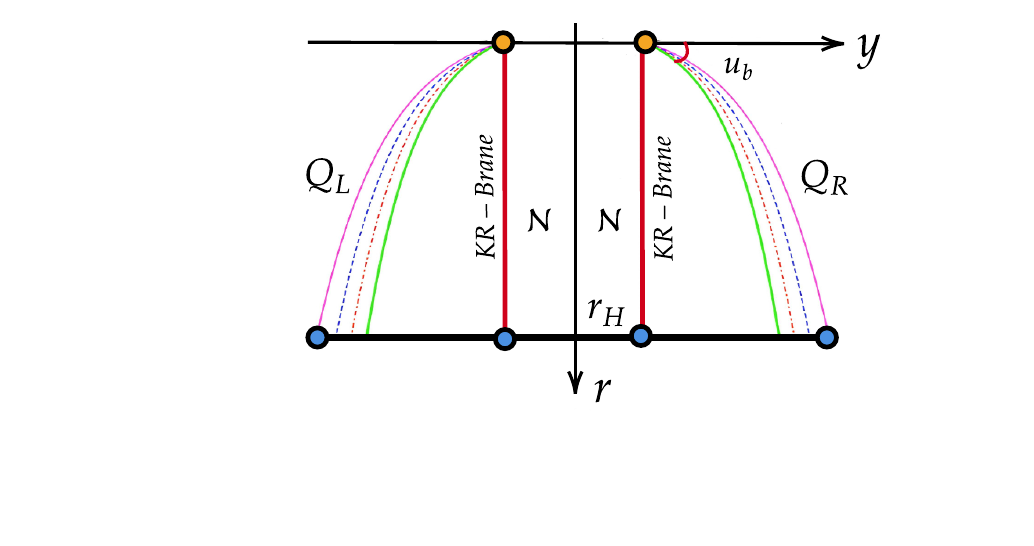}
\vspace{-2.8cm}
\caption{{\it{$Q$ boundary profile for the AdS$_3$ black hole within Horndeski 
gravity 
considering the values $\theta'=2\pi/3$, $u_b=\pi-\theta'$, $\alpha=-8/3$, 
$\Lambda=-3$, and with $\gamma=-0.1$ ({\sl solid}), $\gamma=-0.2$ ({\sl 
dashed}), 
$\gamma=-0.3$ ({\sl dot dashed}), and $\gamma=-0.4$ ({\sl thick}). The dashed 
parallel vertical lines represent the ultra-violet (UV) solution  \eqref{19}, 
while the region 
between curve $Q$ negative and positive branches represents the bulk 
$\mathcal{N}$ \cite{Santos:2021orr}.}}}
\label{profile}
\end{center}
\end{figure}

In  Fig. 
\ref{profile} we depict the $Q$ boundary profile for the AdS$_3$ black hole 
within Horndeski gravity. As we observe, the internal dynamics of the KR-braneworld  
and the presence of Horndeski gravity complicates the  
behavior of the entanglement islands  \cite{Brito:2018pwe,Santos:2023eqp}.
In particular, the formation of the largest 
possible island, starting from a critical anchor and ending at the fault, 
serves as a key feature in determining the entanglement structure of the 
system. The critical anchor represents a   point from which the entanglement island 
can expand, while the fault acts as a boundary restricting this expansion. This 
relation emphasizes the importance of geometric considerations in entanglement 
dynamics, showing how the shapes and connections of these regions directly 
affect the overall entanglement entropy \cite{Geng:2021mic}. Hence,  
atolls are crucial to understand  the phase structure of the system, since 
their formation and characteristics can lead to different phases of entanglement, 
which may be associated with varying behaviors of the entanglement entropy.

\section{Entanglement entropy and Hatman-Maldacena surface}\label{ETER}

In this section  we present the entanglement entropy to the subregion 
$\mathcal{R}$, $\bar{\mathcal{R}}$ and defect, which leads to the island on the brane. In general, it is possible to find 
extremal surfaces $\Omega$ that satisfy the holonomy constraint like 
$\partial\,\Omega=\partial\mathcal{\mathcal{R}}\cup\partial\bar{\mathcal{R}}$.
In scenarios involving defects or branes, the presence of an island can 
complicate the calculation of entanglement entropy. The interaction between the 
two islands is essential for ensuring that the entanglement entropy adheres to unitarity. As the system 
evolves, the total entropy reflects information conservation. The dominance of 
an island surface over an Hatman-Maldacena (HM) \cite{Hartman:2013qma}  surface 
at an initial time indicates that the entanglement dynamics can shift, revealing  the 
non-trivial evolution of entanglement in these scenarios. Page time is particularly 
relevant here, as it marks the moment when the entanglement entropy transit from 
being dominated by the HM surface to the island surface.   The dependence of Page 
time on parameters such as $u_b$, $\alpha$, $\gamma$, and the sizes $r_L$ and $r_R$, indicates the 
richness of the model. Each parameter plays a role in determining the degrees 
of freedom in the defect and the size of the regions of interest \cite{Takayanagi:2011zk}. 
Finally, the observation that Page time is approximately proportional to the 
difference in area between the RT surfaces at $t=0$, reveals a geometric aspect 
of the entanglement dynamics, and this area difference reflects the competing 
contributions to the entanglement entropy and significantly determines how the 
entropy evolves.

Before presenting the entanglement entropy and Hatman-Maldacena surface in Horndeski's scenario. We will present entanglement entropy calculation in the Karch-Randall braneworld in Sec. \ref{KHEE}, it emerges naturally as presented by \cite{Geng:2024xpj}. Based on this scenario, we can extend the calculation of entanglement entropy to the Horndeski scenario Sec. \ref{HGEE}.
\subsection{Entanglement entropy calculation in the Karch-Randall braneworld}\label{KHEE}
The Karch-Randall braneworld scenario provided a natural setting to study Hawking radiation from a black hole using holographic tools \cite{Randall:1999vf,Karch:2000ct,Brito:2018pwe,Santos:2023eqp}. However, quantum effects were an open problem until recently, leaving a gap that needs to be filled. This gap was filled in the work \cite{Geng:2024xpj} where the ambient space is (2+1)-dimensional, for which explicit calculations can be made in each configuration description. 

As discussed and shown by \cite{Geng:2024xpj}, the Karch-Randall braneworld model involves embedding a subcritical brane within an ambient asymptotically  AdS$_{d+1}$ spacetime, denoted as $\mathcal{N}_{d+1}$. In this framework, the ambient spacetime is governed by classical Einstein gravity, while the brane is a codimension-one hypersurface, $\mathcal{N}_{d}$. The action governing this system is given by
\begin{equation}
S=-\frac{1}{16\pi G_{d+1}}\int_{\mathcal{N}_{d+1}} d^{d+1}x\sqrt{-g} (R-2\Lambda)-\frac{1}{8\pi G_{d+1}}\int_{\mathcal{N}_{d}}d^{d}x\sqrt{-h}(K-T)\,,\label{eq:action1}
\end{equation}
The boundary conditions for the bulk metric in the Karch-Randall braneworld model are of the Neumann type near the brane and the Dirichlet type near the remaining asymptotic boundary \cite{Geng:2023qwm,Geng:2021mic}. This implies nonzero metric fluctuations near the brane, while fluctuations vanish near the asymptotic boundary. In this framework, the brane represents a gravitational AdS$_{d+1}$ spacetime, whereas the asymptotic boundary acts as a $d$-dimensional nongravitational bath \cite{Geng:2021iyq,Geng:2020fxl}.

Two sets of equations of motion govern the system:

\begin{itemize}
    \item Einstein's equations, which ensure the vanishing of the bulk variation of the action (\ref{eq:action1}), and the brane embedding equations, vanishing to the brane for the variation of the same action;
    \item Einstein's equations determine the background geometry of the ambient spacetime. In contrast, the brane embedding equations dictate how the brane is embedded within this spacetime, thereby defining the induced geometry on the brane \cite{Santos:2021orr,Geng:2024xpj}.
\end{itemize}

\begin{figure}[!ht]
\begin{center}
\includegraphics[scale=0.65]{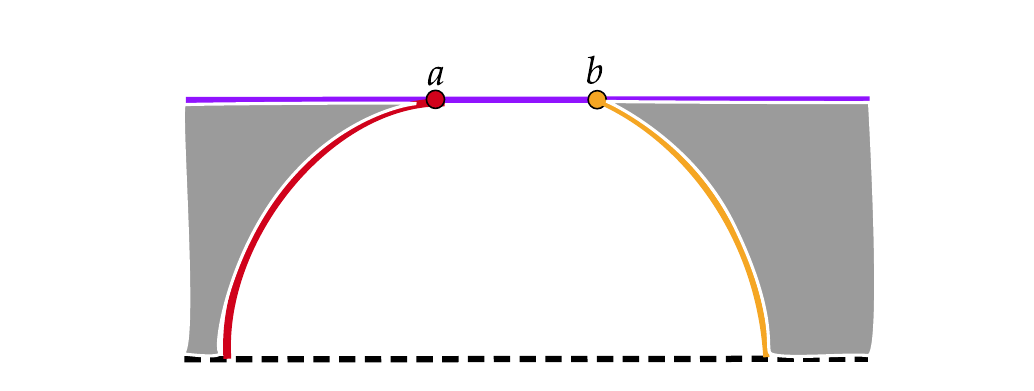}
\end{center}
\caption{{\it{Configuration with two branes in the BTZ background where the gray regions represent cut off by the branes.}}}\label{GAOH0}
\end{figure}

For the bulk description is asymptotically AdS$_3$ with Karch-Randall braneworld \cite{Karch:2000ct}, this description is associated to BCFT$_2$; we want to calculate the subregion entanglement entropy S$_R$ of the thermal field dual state \cite{Geng:2021mic}. To study the thermal field double states $|TFD\rangle$, one intrudes two Karch-Randall branes (see Fig. \ref{GAOH0}) into the spacetime of the ambient BTZ black hole; the BCFT$_2$ are in this double state. In this boundary description, the two BCFT$_2$ live in strips with boundary conditions corresponding to Cardy states \cite{Cardy:2004hm}.

\begin{figure}[!ht]
\begin{center}
\includegraphics[scale=0.65]{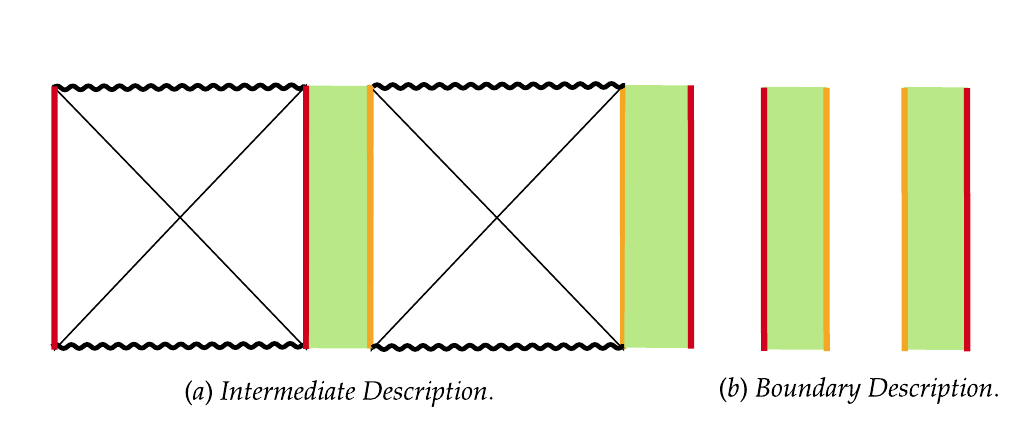}
\end{center}
\caption{{\it{The figure shows in (a) Penrose diagram of the intermediate description for two branes at the bottom of the BTZ black hole; we have two ($1+1$)-dimensional black holes, which are coupled to each other through strip-shaped baths (the green shaded region). We identify the two outer edges in red. It can be seen that the black hole singularities are orbifold singularities inherited from the BTZ black holes. In part of frame (b), we have a description of the boundary of the configuration with two branes at the bottom of the BTZ black hole.}}}\label{GAOH}
\end{figure}

In the setup illustrated in Fig. \ref{GAOH} \cite{Geng:2024xpj}, we consider a model involving two eternal black holes in (1+1)-dimensions, which are coupled through a thermal bath (see Fig. \ref{GAOH}-{\color{blue}a}). This configuration is analyzed within the framework of Boundary Conformal Field Theory (BCFT), where the boundaries are characterized by specific boundary conditions associated with Cardy states (refer to Fig. \ref{GAOH}-{\color{blue}b}) \cite{Ryu:2006bv,Cardy:2004hm,Geng:2024xpj}. This setup provides a novel perspective on the interaction dynamics between black holes and their environments, offering insights into the underlying quantum gravitational processes.

\begin{figure}[!ht]
\begin{center}
\includegraphics[scale=0.65]{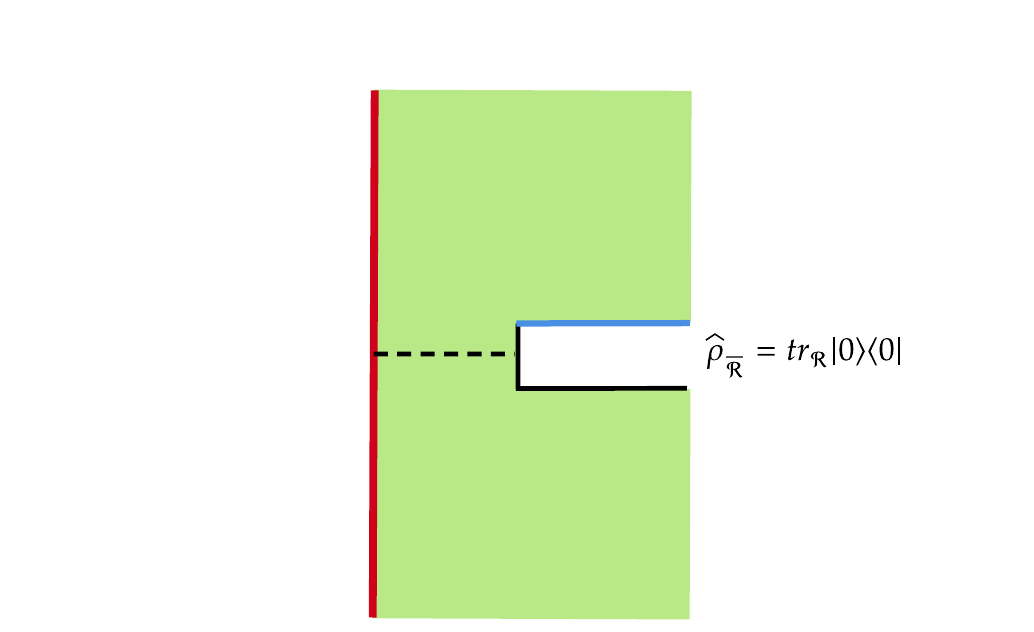}
\caption{{\it{The diagram illustrates the Euclidean path integral formulation with an initially prepared ground state, denoted as $|0\rangle$. Within the dashed interval, we identify the subregion $\mathcal{R}$. The reduced density matrix operator for this subregion, $\hat{\rho}_{\mathcal{R}}$, is derived from the full density matrix $\hat{\rho}=|0\rangle \langle 0|$ by integrating over all possible field configurations outside of $\mathcal{R}$. This approach allows us to isolate the contributions specific to the subregion, providing insights into the entanglement properties and quantum correlations present in the system.}}}\label{GAOH1}
\end{center}
\end{figure}

Through the holographic dictionary of the Karch-Randall braneworld, one can perform a mapping of the replicated boundary partition function into a Euclidean gravitational path integral as follows:

\begin{equation}
    Z_{\text{BCFT}}[\mathcal{N}^{(n)}]=\int_{\partial\mathcal{N}^{(n)}_{d+1}=\mathcal{N}^{(n)}} D[g] e^{-S}=Z_{\text{grav}}[\mathcal{N}^{(n)}]\,,
\end{equation}
Here $S$ is the gravitational action Eq. (\ref{eq:action1}) where $\mathcal{N}^{(n)}_{d+1}$ is the manifold in the bulk as the $n$-branched half-plane with the branching point at the coboundary of $\mathcal{R}$ and $\mathcal{\bar{R}}$; it must have a conformal boundary as the replica manifold $\mathcal{N}^{(n)}$ together with Karch-Randall branes $\mathcal{N}^{(n)}_{d}$ whose asymptotic boundary $\partial\mathcal{N}_{d}$ is the same as $\partial\mathcal{N}^{(n)}$ \cite{Geng:2023qwm}. Thus, we are integrating over smooth metric configurations. Then, the gravitational path integral can be computed using the saddle point approximation; the gravitational theory in the bulk is classical \cite{Basu:2023jtf}. The entanglement entropy can be computed as
\begin{eqnarray}
S_{\mathcal{R}}&=&-tr\hat{\rho}_{\mathcal{\bar{R}}}\log\hat{\rho}_{\mathcal{\bar{R}}}=\lim_{n\rightarrow1}\frac{1}{1-n}\log\,tr \hat{\rho}_{\mathcal{\bar{R}}}^{n}\,,\\
    &=&\lim_{n\rightarrow1}\frac{1}{1-n}\frac{Z_{\text{grav}}[\mathcal{N}^{(n)}]}{Z_{\text{grav}}[\mathcal{N}^{(0)}]^{n}}\,,
\end{eqnarray}
and considering two Karch-Randall branes in the ambient BTZ black hole spacetime Fig \ref{GAOH1}. To calculate the entanglement entropy between the subregion $\mathcal{R}$ and its complement $\mathcal{\bar{R}}$, the reduced density matrix operator is used 
\begin{equation}
\hat{\rho }_{\mathcal{\bar{R}}} =tr_{\mathcal{R}}|0\rangle \langle 0|\,.\label{eq:key}
\end{equation}

Through the half Euclidean plane with the Cardy boundary-$\mathcal{N}$ \cite{Ryu:2006bv,Cardy:2004hm,Geng:2024xpj}, the trace of the $n-th$ power of the normalized reduced density matrix operator is given by
\begin{equation}
    tr \hat{\rho}_{\bar{R}}^{n}=\frac{Z_{BCFT}[\mathcal{N}^{(n)}]}{Z^{n}_{BCFT}[\mathcal{N}]}\,.\label{eq:key}
\end{equation}
This expression is the key point to prove the Ryu-Takayanagi \cite{Ryu:2006bv} conjecture in the AdS/BCFT correspondence scenario using gravitational path integral. Thus, with this, we can conclude that
\begin{eqnarray}
S_{R}&=&\lim_{n\rightarrow1}\frac{1}{1-n}\frac{e^{-nS[g^{BTZ}]+\frac{A(\gamma)}{4G_{d+1}}\frac{1-n}{n}]}}{e^{-nS[g^{BTZ}]}}\,,\\
    &=&\frac{A(\gamma)}{4G_{d+1}}\,.
\end{eqnarray}
where $g^{EBTZ}$ is the Euclidean BTZ metric. The phases of the Karch-Randall brane within the manifold $\mathcal{N}_{d+1}^{(n)}$ \cite{Geng:2021mic,Geng:2022dua}. 

With this prescription, we can identify two distinct phases: 
\begin{itemize}
    \item First phase: characterized by $n$-disconnected components and another by a single component that connects the Cardy boundaries of the boundary replica manifold $\mathcal{N}_{d+1}^{(n)}$ \cite{Cardy:2004hm}. In this phase, the conical singularity $\gamma$ remains in the bulk, linking the two $\partial\mathcal{R}$ boundaries;
    \item Second phase: $\gamma$ extends to the brane, with $\mathbb{Z}_{n}$  fixed points present, resulting in two disconnected components \cite{Tonni:2010pv} that connect the boundaries of $\partial\mathcal{R}$ to the brane \cite{Geng:2023qwm,Geng:2021hlu}.
\end{itemize}
This dual-phase behavior provides new insights into the geometric and topological properties of the Karch-Randall brane, offering potential implications for understanding brane dynamics in higher-dimensional spaces \cite{Geng:2022dua}. However, we have that
\begin{equation}
    S(R)=\min (\frac{A_{c}}{4G_{d+1}},\frac{2A_{dc}}{4G_{d+1}})\,,\label{eq:RTderiation}
\end{equation}
This is the Ryu-Takayanagi conjecture \cite{Ryu:2006bv}. Note that $\gamma_{dc}$ must also have its area minimized over its possible endpoints on the brane; this corroborates our saddle point approximation. This discussion shows that the Ryu-Takayanagi conjecture is proven; one can see that, indeed, the emergence of the disconnected Ryu-Takayanagi surface is due to the connected phase of the Karch-Randall brane in the replica path integral \cite{Basu:2023jtf}.

\subsection{Derivation of the functional entanglement entropy for Horndeski gravity }\label{HGEE}
We must derive the functional area for Horndeski gravity's entanglement entropy to study the holographic entanglement entropy. Considering the Horndeski Lagrangian ${\cal L}_H$, and following the technique detailed in previous works of \cite{DosSantos:2022exb,Caceres:2017lbr,Feng:2015oea,Wald:1993nt,Iyer:1994ys}, which consider:
\begin{itemize}
    \item The Lagrangian-${\cal L}_H$ of Horndeski theory does not include quadratic or higher-order curvature terms;
    \item Incorporates a Wald-type term \cite{Feng:2015oea}, which is important in developing holographic entanglement entropy in arbitrary gravity theories.
\end{itemize}
This approach facilitates the application of the replica trick, a method essential for deriving holographic entanglement entropy. Through the microscopic definition of entanglement entropy \cite{Ryu:2006ef,Caceres:2017lbr}:

\begin{eqnarray}
S_{EE}=-Tr(\rho\log(\rho))
\end{eqnarray}
 Here, $\rho$ is the reduced density matrix of a subregion of a time slice of the boundary. Here, the idea is to find the entanglement entropy in the subregion 
$\mathcal{R}$, $\bar{\mathcal{R}}$. For this, we apply the replica trick through the $n^{th}$ R\' {e}nyi entropy, defined by:
\begin{equation}
    S_{n} = -\frac{1}{n-1} \log{\mathrm{Tr}{(\rho^{n})}}.
\end{equation}
Here the entanglement entropy is the analytical continuation of $S_{n}$ as $n \rightarrow 1$: $S_{EE} = \lim_{n \rightarrow 1} S_{n}$. From the overview point of path integral, $\rho$, can be express $S_{n}$ in terms of the partition function $Z_{n}$ considering an appropriate n-sheeted Riemann surface $M_{n}$ as:
\begin{equation}\label{SnZn}
    S_{n} = -\frac{1}{n-1} \left( \log{Z_{n}} - n\log{Z_{1}} \right),
\end{equation}
in the above equation $Z_{1}$ is the original partition function. Besides, using the basic holographic relation $Z_{CFT} = e^{-S_{bulk}}$ between the field theory partition function $Z_{CFT}$ and the bulk action for an appropriate bulk geometry. With this and through the prescription of \cite{Caceres:2017lbr,Feng:2015oea}, we can derive the area functional for Entanglement Entropy for Horndeski gravity ($S_{EEH}$):
\begin{equation}
S_{EEH}=\int_{\mathcal{N}}{d^{d}x\sqrt{\eta}\frac{\partial{\cal L}_H}{\partial R_{\mu\nu\rho\sigma}}\epsilon_{\mu\nu}\epsilon_{\rho\sigma}},
\end{equation}
where $\epsilon_{\mu\nu}$ and $\epsilon_{\rho\sigma}$ denotes the bi-normal to $\mathcal{N}$ \cite{Feng:2015oea,Wald:1993nt,Iyer:1994ys}. We can derive Horndeski Lagrangian with respect to the Riemann tensor as:

 \begin{eqnarray}\label{dLdR}
    \frac{\partial{\cal L}_H}{\partial R_{\mu\nu\rho\sigma}} &=& - \frac{\kappa}{32\pi} (g^{\mu\rho}g^{\nu\sigma} - g^{\nu\rho}g^{\mu\sigma}) - \frac{\gamma}{128\pi}[ g^{\mu\rho}\phi^{,\nu}\phi^{,\sigma} - g^{\nu\rho}\phi^{,\mu}\phi^{,\sigma} + g^{\nu\sigma}\phi^{,\mu}\phi^{,\rho} - g^{\mu\sigma}\phi^{,\nu}\phi^{,\rho} \nonumber \\
    &-& (g^{\mu\rho}g^{\nu\sigma} - g^{\nu\rho}g^{\mu\sigma}) \phi^{,\lambda} \phi_{,\lambda}]
\end{eqnarray}
Here in our work $\phi^{,\alpha} = g^{\alpha\beta} \phi_{,\beta}$. For the metric in complex coordinates like:
\begin{equation}
ds^{2} = dzd\bar{z} +\eta_{ij} dy^{i}dy^{j}
\end{equation}
Upon substituting the metric components into (\ref{dLdR}), we have
\begin{equation}
\frac{\partial{\cal L}_H}{\partial R_{z\bar{z}z\bar{z}}} = \frac{\kappa}{8\pi} - \frac{\gamma}{32\pi} \left( \phi^{,\lambda} \phi_{,\lambda} - \phi^{,z} \phi^{\bar{,z}} \right)
\end{equation}
Upon further expanding $\phi^{,\lambda}\phi_{,\lambda} = \phi^{,z}\phi^{,\bar{z}} + \eta_{ij}\phi^{,i}\phi^{,j}$, this further simplifies to:
\begin{equation}
\frac{\partial{\cal L}_H}{\partial R_{z\bar{z}z\bar{z}}} = \frac{\kappa}{8\pi} - \frac{\gamma}{32\pi} \eta_{ij} \phi^{,i} \phi^{,j}
\end{equation}
Thus, we have that the functional for holographic entanglement entropy then read \cite{Caceres:2017lbr,Feng:2015oea}:
\begin{equation}
S_{EEH}=\int{d^{d}x\sqrt{\eta}\frac{\partial{\cal L}_H}{\partial R_{z\bar{z}z\bar{z}}}}=\frac{\kappa}{4} \int{d^{d}x\sqrt{\eta}\left[1-\frac{\gamma}{4 \kappa} \eta_{ij} \phi^{,i} \phi^{,j} \right]}\label{EEG}
\end{equation}
 We can note that comparing our result with the Ryu-Takayanagi formula \cite{Ryu:2006ef}, the correction for the area is due to the term $\gamma\,G_{\mu\nu}\nabla^{\mu}\phi\nabla^{\nu}\phi$ where for $\gamma=0$ we recover the usual Einstein case. Considering the metric (\ref{me}) and (\ref{EEG}), we can derive the area functional $\mathcal{A}(\partial\mathcal{\mathcal{R}}\cup\partial\bar{\mathcal{R}})$ to the subregion $\mathcal{R}$ and $\bar{\mathcal{R}}$ for Horndeski gravity, which is given by
\begin{eqnarray} 
&&S_{EEH}=\frac{\mathcal{A}}{4G}\\
&&\mathcal{A}=\int^{\pi}_{u_{b}}{du\frac{\chi}{r^2\sin^2(u)}\sqrt{r^2+\frac{r'^{
2}(u)}{f(r)}}},\label{Entangle}\\
&&\chi=1+\frac{3(\alpha+\gamma\Lambda)}{\alpha\,A(u)}.
\end{eqnarray}
Now, the Euler-Lagrange equation for the action (\ref{Entangle}) reads
\begin{eqnarray}
r{''}=-rf(r)+2\cot(u)r{'}+\frac{2\cot(u)r{'3}}{r^2f(r)}-\frac{r{'}\chi{'}}{\chi}
-\frac{r{'3}\chi{'}}{r^2f(r)\chi}+\frac{r{'2}}{r}.
\end{eqnarray}
However, through Fig. \ref{p2} and using trigonometric identities  we can see that $\cos[2(u\to\pm\infty)]\to1$ and $A(u)\to\,2(\alpha-3\gamma)$, which provide 
\begin{eqnarray}
\chi=\frac{2-\beta_0(1-\beta_0)}{2(1-\beta_0)}=constant.
\end{eqnarray}
Hence, we recover the result of \cite{Geng:2021mic}, namely   
\begin{eqnarray}
r{''}=-rf(r)+2\cot(u)r{'}+\frac{2\cot(u)r{'3}}{r^2f(r)}+\frac{r{'2}}{r}.
\end{eqnarray}
\begin{figure}[!ht]
\begin{center}
\includegraphics[scale=0.75]{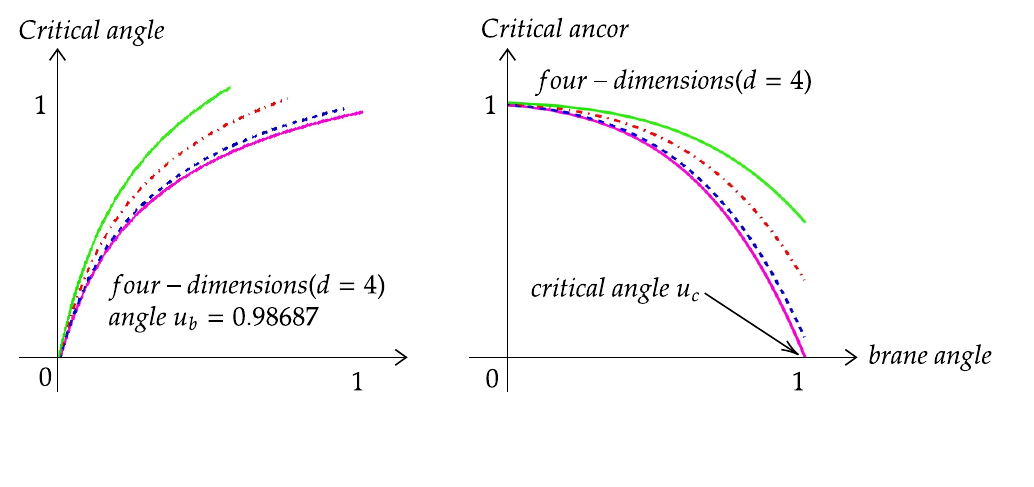}
\vspace{-1.cm}
\caption{{\it{The   critical angle and ancor with $\alpha=-8/3$ and 
$\Lambda=-3$ for different values of the  Horndeski parameter $\gamma$, namely  $\gamma=-0.1$ ({\sl solid}), $\gamma=-0.2$ ({\sl dashed}), $\gamma=-0.3$ ({\sl dot dashed}), and $\gamma=-0.4$ ({\sl thick}). The critical angle is a monotonically increasing function of the number of spatial dimensions $d=4$, thus the behavior of the critical anchor that defines the beginning of the atoll for the black string also assumes different values (see Fig. \ref{ATOLL}). Finally, the critical 
anchor increases monotonically with the brane angle and decreases as it 
coincides with the defect at the critical angle.}}}
\label{ancor}
\end{center}
\end{figure}

As we  can see,  according to \cite{Geng:2021mic}  the boundary terms in the 
variation of $\mathcal{A}$
vanish  by imposing boundary conditions on $\Sigma$. This occurs since a 
Dirichlet condition is imposed on the conformal boundary and a Neumann 
condition 
\cite{Geng:2020fxl}. However, the angles are
deformed by the Horndeski parameters and must anchor on the brane at right 
angles; they will be right angles only for $\alpha=0$ and $\gamma=0$. Thus, the 
Euler-Lagrange equation with Horndeski parameters $\alpha$ and $\gamma$ leads 
to nontrivial physics below the critical angle we explore. 
In Fig. \ref{ancor} we analyze how the critical angle depends on $\alpha$ and 
$\gamma$, giving the trajectories of RT surfaces becoming circular geodesics 
that exist for all brane angles.

\subsection{Hartman-Maldacena entanglement entropy} 

We use the spatial embedding formalism to compute the Hartman-Maldacena (HM)
entanglement entropy \cite{Hartman:2013qma}.
Thus, the geometry is embedded as a codimension-one sub-manifold of a 
four-dimensional Minkowski space:
\begin{eqnarray}
ds^2=\eta_{ab}dX^{a}dX^{b};\,\eta_{ab}=diag(1,1,-1,-1),
\end{eqnarray}
where the embedding equation is $X_{a}X^{a}=1$. For convenience, we impose the  
 re-parameterization $sin^{-1}(u)=\cosh(\rho)$ for the metric (\ref{me}), and 
considering the fact that a dual field theory is a BCFT$_2$ on an AdS$_2$ black 
hole 
background with conformal boundary conditions at $r=0$, we obtain
\begin{equation}
ds^{2}_{AdS_4}=\frac{\cosh(\rho)}{r^{2}}\left(-f(r)dt^{2}+\frac{dr^{2}}{f(r)}
\right)+d\rho^2.\label{med1}
\end{equation}
To simplify our results, we rewrite $f(r)$ as 
\begin{equation}
f(r)=\omega\left(1-\frac{r^2}{r^{2}_{h}}\right);\,\omega=1+\left(\frac{
\alpha-3\gamma}{\gamma}\right)\cosh^2(\rho),\label{med2}
\end{equation}
and therefore the metric (\ref{med1}) can be recovered using the following 
parameterization of the embedding equation
\begin{eqnarray}
&&X_{0}=\frac{2r_h-r}{r}\cosh(\rho)\\
&&X_{1}=\frac{2r_h}{r}\sqrt{\omega\left(1-\frac{r^2}{r^{2}_{h}}\right)}
\sin\left(\frac{2\pi\,t}{\beta}\right)\cosh(\rho)\\
&&X_{2}=\frac{2r_h}{r}\sqrt{\omega\left(1-\frac{r^2}{r^{2}_{h}}\right)}
\cos\left(\frac{2\pi\,t}{\beta}\right)\cosh(\rho)\\
&&X_{3}=\sinh(\rho),
\end{eqnarray}
with $\beta=4\pi\,r_h=2\omega/T$   the inverse Hawking temperature. 

The idea behind the spatial embedding formalism is that we can calculate the 
Hartman-Maldacena surface area without solving the minimum area (geodesic) 
differential equation. Using this embedding, the length $l$ can be computed 
with 
the following coordinates ($X_0,\,X_1,\,X_2,\,X_3$) and 
($X'_0,\,X'_1,\,X'_2,\,X'_3$): 
\begin{eqnarray}
l=\cosh^{-1}(X_0X'_0X_1X'_1-X_2X'_2-X_3X'_3).
\end{eqnarray}
As shown in the schematic diagram of Fig. \ref{p1}, while the island surface 
begins at the bipartition and ends at the KR brane, we can observe that the HM 
surface \cite{Hartman:2013qma} passes through an Einstein bridge-Rosen and ends 
at the right bipartition in the double thermal field on the right side. In our 
diagram, the left and right bipartitions are located at $(u; \rho) = (u_L; 1)$ 
and $(u; \rho) = (u_R; 1)$. As discussed in \cite{Geng:2022dua}, one can 
introduce a regularization parameter $\rho_\epsilon$, which can be set to 1; 
our bipartitions are on the asymptotic boundary. 
We can note that in order to obtain a correct bipartition we must take the time 
coordinate $t\to-t+\frac{i\beta}{2}$. This corresponds to the reversal of the 
Killing vector field similarly to the time on the other side of the black 
string 
horizon \cite{Geng:2022dua} (see Fig. \ref{p1}).

We proceed by introducing the quantities
\begin{eqnarray}
\Delta_L=r_h-r_L,\,\Delta_R=r_h-r_R.
\end{eqnarray}
Hence, the  HM surface can be written as
\begin{eqnarray}
&&\!\!\!\!\!\!\!\!\!\!\!\!\!\!\! \!
\mathcal{A}_{HM}
=\chi\cosh^{-1}
(-X^L_0X^R_0-X^L_1X^R_1+X^L_2X^R_2+X^L_3X^R_3)\nonumber\\                
\!\!\!\!\!\!\!\!\!\!\!&&=\chi\cosh^{-1}\left[\frac{
(2r_h-r_L)(2r_h-r_R)+4r_h\omega(\rho_ { \epsilon}
)\sqrt{\Delta_L\,\Delta_R}\,\cosh\left(\frac{2\pi\,t}{\beta 
}\right)}{r_Lr_R}\cosh(\rho_{\epsilon })-\sinh^2(\rho_{\epsilon 
})\right]\!\!.
\end{eqnarray}
 Using the hyperbolic trigonometric identities, we acquire
\begin{eqnarray}
S_{HM}&=&\frac{\mathcal{A}_{HM}}{4G}\\      
&=&\frac{c}{6}log\,\left[\frac{r_h}{r_Lr_R}\left(\Delta_L+\Delta_R+2\omega(\rho_
{\epsilon})\sqrt{\Delta_L\,\Delta_R}\,\cosh\left(\frac{4\pi\,t}{\beta 
}\right)\right)\right]+\frac{c}{3}\rho_{\epsilon}\label{HME},
\end{eqnarray}
where $c=3\chi/2G$ is the central charge-like for Horndeski gravity 
\cite{DosSantos:2022exb}.
Thus, considering a particular case with $r=r_L=r_R$, and using 
$\Delta_r=r_h-r$, we have
\begin{eqnarray}
S_{HM}=\frac{c}{6}log\,\left[\frac{2r_h\Delta_r}{r^2}\left(1+\omega(\rho_{
\epsilon})\sqrt{\Delta_L\,\Delta_R}\,\cosh\left(\frac{4\pi\,t}{\beta 
}\right)\right)\right]+\frac{c}{3}\rho_{\epsilon}.
\end{eqnarray}

We can now obtain the result with the pair of minimal island surfaces that 
cross from the bipartitions $r_L$ and $r_R$ to the corresponding physical branes 
location at 
$\rho=\rho_{*}$. In 
particular, 
we can schematically represent $\mathcal{A}_{island}$ as:
\begin{eqnarray}
&&\mathcal{A}_{island} =\chi\int_{min.\ island} ds\ ;\ ds=d\rho 
\sqrt{1+\frac{cosh^{2}( \rho )}{r^{2}( \rho ) \omega \left( 1-\frac{r^{2}( \rho 
)}{r_{h}}\right)} r^{\prime 2}( \rho )}\\
&&r^{\prime 2}( \rho )\rightarrow \mathcal{A}_{island} =2\chi\int^{\rho }_{\rho 
_{*}}{d\rho}=2( \rho -\rho _{*}).
\end{eqnarray}
 In this 
case,  the entanglement entropy calculated by the island surface is given by
\begin{eqnarray}
S_{island}=\frac{\mathcal{A}_{island}}{4G}=-\frac{c}{3}\rho_{*}+\frac{c}{3}\rho_
{\epsilon},
\end{eqnarray}
where the prescriptions of RT \cite{Ryu:2006bv}  show that the entanglement 
entropy is the minimum of $S_{HM}$ and $S_{island}$, i.e.  
$S=min(S_{HM},S_{island})$.

\subsection{Field theory calculation} 

In this subsection we calculate the entanglement entropy between 
subsystem $A$ and its complement. This calculation is equivalent to calculating a two-point function 
$\langle\Phi_n(r_R,t)\,\Phi_n(r_L,t)\rangle$ of the twist operator fields 
$\Phi_n(r,t)$ inserted in the two bipartition points $r_L$ and $r_R$. Hence, 
since the field theory lives on a curved background, in order to perform the 
entanglement entropy calculation on the field theory side  we need the appropriate geometry 
  \cite{Geng:2022dua,Doroudiani:2019llj}.

\begin{figure}[!ht]
\begin{center}
\includegraphics[scale=0.5]{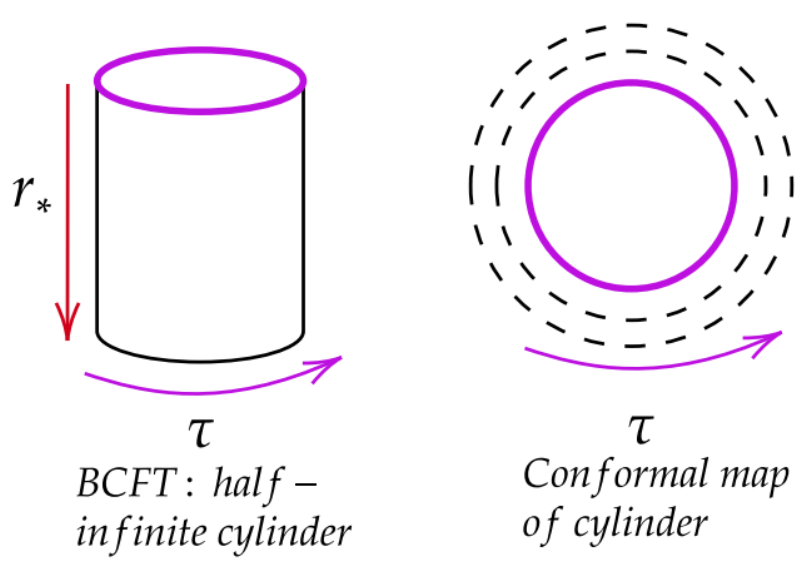}
\caption{ {\it{Schematic map of a field theory on the curved geometry 
of a flat background, using  the conformal transformations.}}}
\label{Confro}
\end{center}
\end{figure}

We can calculate the entanglement entropy of subsystem $A$ at the boundary
($S_{bdry}$)
 and 
in the bulk ($S_{bulk}$) channels, using the 
conformal transformations \cite{Basu:2023jtf,Geng:2022dua}:
\begin{eqnarray}
&&ds^{2} =\frac{1}{r^{2}}\left[ -\omega \left( 1-\frac{r}{r_{h}}\right) dt^{2} 
+\frac{dr^{2}}{\omega \left( 1-\frac{r}{r_{h}}\right)}\right]\\
&&ds_{conformal}^{2} =\Omega ^{2}( r_{*})\left( -dt^{2} +dr^{2}\right)\\
&&\Omega ^{2}( r_{*}) =\frac{1}{r}\sqrt{\omega \left( 
1-\frac{r}{r_{h}}\right)}\\
&&r_{*} =-r_{h} log\left[ \omega \left( 1-\frac{r}{r_{h}}\right)\right],
\end{eqnarray}
with Conformal boundary (see Fig. \ref{Confro})
\begin{eqnarray}
&&z=r_{*} +i\tau \ ;\ \overline{z} =r_{*} -i\tau\\
&&ds^{2} =\Omega ^{2}( r_{*}) dzd\overline{z} =\Omega ^{2}( r_{*}) 
e^{-\frac{r_{*}}{r_{H}}} d\sigma d\overline{\sigma }.
\end{eqnarray}
The twist fields $\Phi_n$ inserted at the two bipartition point 
$r_L$ and $r_R$ satisfy 
\begin{eqnarray}
S_A=\lim_{n\to\,1}\frac{1}{1-n}log(\langle\Phi_n(r_R,t_R)\,\Phi_n(r_L,
t_L)\rangle).
\end{eqnarray}
Finally, we find
\begin{eqnarray}
&&S_{bdry}=2\,log(g_b)+\frac{c}{3}\log\left(\frac{2}{\epsilon}\right)\\
&&S_{bulk}=\frac{c}{6}log\,\left[\frac{r_h}{r_Lr_R}
\left(\Delta_L+\Delta_R+2\omega(\rho_{\epsilon})\sqrt{\Delta_L\,\Delta_R}\,
\cosh\left(\frac{4\pi\,t}{\beta 
}\right)\right)\right]+\frac{c}{3}\log\left(\frac{2}{\epsilon}\right),
\end{eqnarray}
 with $\Delta_L=r_h-r_L$, $\Delta_R=r_h-r_R$
and where $\epsilon$ is the UV cutoff.

\begin{figure}[!ht]
\begin{center}
\includegraphics[scale=0.85]{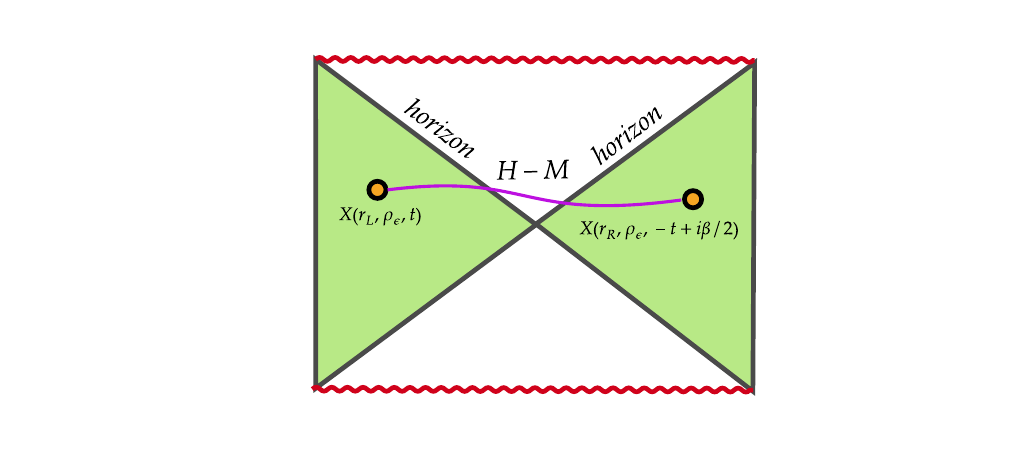}
\vspace{-1.3cm}
\end{center}
\caption{{\it{Schematic form of Hartman-Maldacena (HM) surface connecting 
boundary field theory, which consists of two copies of a holographic CFT with BCFT boundary. In 
the right bipartition the time coordinate $t\to-t+i\beta/2$ was considered, 
which corresponds to a reversal of the time-like Killing vector field on the other 
side of the horizon of the black string.}}}\label{PHASE1}
\end{figure}

According to \cite{Santos:2021orr}, we can 
calculate the Horndeski gravity boundary entropy      of the boundaries 
$Q_{L}$ 
and $Q_{R}$ for the two bipartitions as
\begin{eqnarray}
&&S^{L}_{bdry}=\frac{2\,\Delta 
y_{Q_L}}{3r_{L}}\left(1-\frac{\xi}{8}\right)-\frac{\xi\,b(u_b)}{3r^{2}_{L}}
\left(1-\frac{\xi}{4}\right)-\frac{\xi\,h(u_b)\cot(u_b)}{r^{2}_{L}}+\frac{\xi\,
q(u_b)}{3}\\
&&S^{R}_{bdry}=\frac{2\,\Delta 
y_{Q_R}}{3r_{R}}\left(1-\frac{\xi}{8}\right)-\frac{\xi\,b(u_b)}{3r^{2}_{R}}
\left(1-\frac{\xi}{4}\right)-\frac{\xi\,h(u_b)\cot(u_b)}{r^{2}_{R}}+\frac{\xi\,
q(u_b)}{3},
\end{eqnarray}
with
\begin{eqnarray}\label{btheta}
b(u_b)&=&\cos(u_b)\tan^{-1}\left(\frac{1}{\sin(u_b)}\right)+\cot(u_b)\left(\frac
{1+\cos^{2}(u_b)\cot^{2}(u_b)}{\sin^{2}(u_b)}\right)\,\\
h(u_b)&=&-\frac{(1+\pi/2)}{2\sin(u_b)}+\frac{\cot^{3}(u_b)\cos^{2}(u_b)}{
(1+\cos^{2}(u_b))}\tanh^{-1}\left(\frac{\sqrt{2}\cos(u_b)}{\sqrt{1+\cos^{2}(u_b)
}}\right)\nonumber\\
&-&\frac{(1+\cos^{2}(u_b)+3\cos^{4}(u_b)-3\cos^{6}(u_b))}{3\sin^{5}(u_b)(1+\cos^
{2}(u_b))}\,, 
\nonumber\\
q(u_b)&=&\left(\frac{1}{4}-\cos^{3}(u_b)\right)\cot(u_b)\csc(u_b)\,. \nonumber
\end{eqnarray}
Hence, if the subsystems   are considered large and far from the boundary then 
we obtain
\begin{eqnarray}
&&S^{L}_{bdry}=\frac{\xi\,q(u_b)}{3}\\
&&S^{R}_{bdry}=\frac{\xi\,q(u_b)}{3} .
\end{eqnarray}
This result expresses a significant fact about the information process in black 
holes (specifically the geometry of the AdS$_4$ black string with a 
Karch-Randall brane \cite{Karch:2000ct,Brito:2018pwe,Santos:2023eqp}). Note 
that this residual information implies that even if the black hole evaporates 
entirely from the point of view of classical entropy, we still have information 
being emitted 
\cite{Susskind:2014rva,Brown:2015bva,Brown:2015lvg,Susskind:2018fmx,
Brown:2018bms,Brown:2017jil,Brown:2019whu,Brown:2022rwi,Santos:2021orr,
Santos:2023mee,Santos:2024zoh}. In this case  $S^{L}_{bdry}$ and 
$S^{R}_{bdry}$ are holographically entangled and are estimated as the minimum 
area \cite{Santos:2021orr,Santos:2023mee,Santos:2024zoh}. Thus, the ``open 
wormhole'' geometry \cite{Takayanagi:2011zk}, if it exists, would be 
responsible 
for the entanglement of $S^{L}_{bdry}$ and $S^{R}_{bdry}$, both connected by a 
 HM surface \cite{Hartman:2013qma} (see Fig. \ref{PHASE1}).

\section{Page curve behavior}\label{T-A-PAGE}

The relation between entropy $S$ and area $A$ in the context of quantum gravity 
\cite{Hawking:1975vcx,Santos:2021orr,Sokoliuk:2022llp,Santos:2023flb,
Santos:2023mee}, particularly with a negative cosmological constant 
\cite{Maldacena:1997re,Witten:1998qj}, indeed connects deeply the holographic 
principle and the Ryu-Takayanagi conjecture \cite{Ryu:2006bv}. Identifying the 
entanglement entropy with the area of minimal surfaces in the bulk serves as a 
bridge between gravitational theories and quantum information theory. This 
conjecture states that the entanglement entropy of a region in a quantum field 
theory is proportional to the area of a minimal surface in the corresponding 
bulk gravitational theory. For a black hole, this translates into the 
understanding that the entropy of the black hole is connected to the area of 
its event horizon. The consideration that the Hilbert space describing black 
hole dynamics is finite-dimensional, composed by orthogonal states often called
``black hole microstates'', is significant 
\cite{Iliesiu:2022kny,Balasubramanian:2022gmo}. 
It suggests that despite the seemingly infinite degrees of freedom in a 
gravitational theory, the effective description can be captured within a finite 
framework, resembling the structure of statistical mechanics. Furthermore, the 
equivalence of quantum gravity with two AdS asymptotic boundaries, to two 
copies of  BCFT, is particularly interesting. Each boundary 
(BCFT$_L$ and BCFT$_R$) can be viewed as encoding information about the 
gravitational system, leading to a richer understanding of entanglement and 
correlation between the two sides. The semiclassical approach to calculate the 
entropy of black holes suggests that one can account for the information 
paradox through the evolution of the Page curve during black hole evaporation. 
This curve reflects how the entanglement entropy changes as the black hole looses 
mass, connecting quantum mechanics, thermodynamics, and information theory. 
Thus,  the semiclassical gravitational 
path integral is not merely a computational tool but offers a convenient 
framework to explore the full Hilbert space dynamics. This perspective may help 
in understanding how information is preserved even when black holes evaporate 
\cite{Balasubramanian:2022gmo}, addressing longstanding concerns about the fate 
of information in gravitational collapse.

The investigation of the Page curve in the context of Horndeski gravity, 
particularly through the AdS/BCFT correspondence, explores black hole 
information. The Page curve describes how a black hole  entanglement entropy $S$ evolves during its 
evaporation. Initially, the entropy increases as the black hole absorbs matter, 
then it decreases as information is emitted. As a subclass of 
scalar-tensor theories, Horndeski gravity includes terms that account for 
higher-order derivatives of the scalar field. This framework allows for a 
richer set of dynamics than standard general relativity, making it relevant for 
understanding modifications to gravitational behavior in the presence of black 
holes. 

The parameters $\alpha$ and $\gamma$ of Horndeski gravity affect the 
dynamics of the black hole and the resulting Page curve, altering spacetime 
geometry and affecting the behavior of the extremal surfaces, which 
contribute to entanglement entropy. Analyzing how these parameters affect the 
Page time (the moment at which the entropy begins to decrease) and Page angle 
(the steepness of the entropy curve during the evaporation process) is critical 
for understanding the interplay between black hole thermodynamics and quantum 
information \cite{Geng:2021mic,Geng:2023qwm,Geng:2021hlu}. In summary, 
exploring the Page curve within Horndeski gravity offers a   framework for 
understanding black hole information dynamics and contributes to the ongoing 
efforts to reconcile quantum mechanics with gravitational phenomena. 
\begin{figure}[!ht]
    \centering
   \includegraphics[scale=0.6]{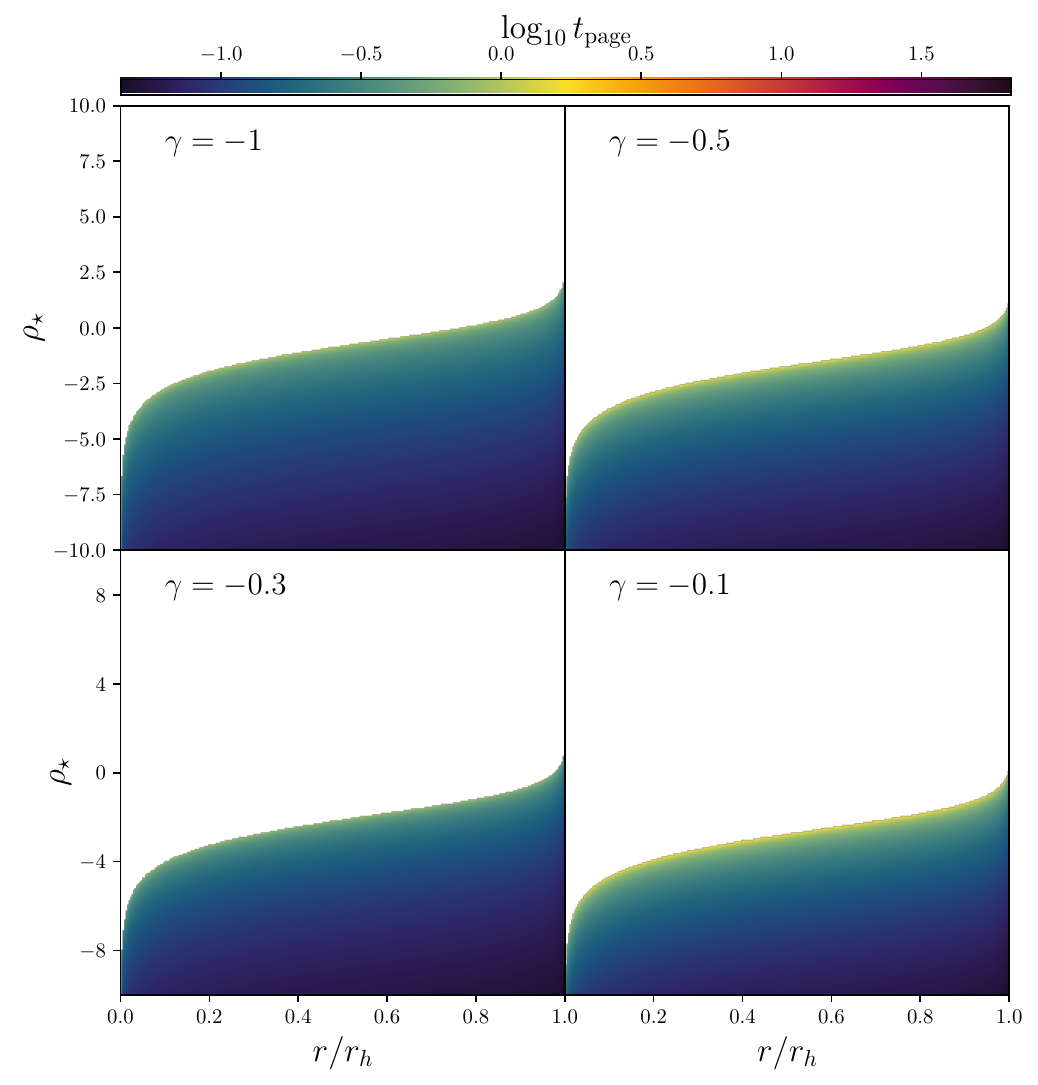}
    \caption{{\it{The Page angle   $\rho_{Page}$ for various values of 
$\gamma$ and fixed $\alpha=-8/3$. The points 
where the Page angle gives the tensionless brane $\rho=0$ are those where the parameter 
$\gamma$ is furthest from zero. The fact that $\gamma$ is far 
from its null value shows that for $\rho_{Page}>0$ the island dominates at $t=0$, 
while for $\rho_{Page}<0$ the HM surface dominates at $t=0$. The
regions showing dominant competition between the island and HM
surfaces initially dominate the entropy calculation due to fixing $\alpha$ 
and varying $\gamma$, and reveal the strength of the scalar field contribution. 
These graphs are equivalent with the density plot along a diagonal slice in 
Fig.  \ref{fig:rho_page_gamma} below, with $r_L=r_R$.}}}
    \label{fig:t_page_gamma}
\end{figure}

We employ a holographic model in AdS$_{3}$ space, using the AdS/BCFT 
correspondence to relate gravitational dynamics to a dual boundary conformal 
field theory. The entanglement entropy is calculated as 
$S=min(S_{HM},\,S_{island})$, where $S_{HM}$ represents the Hartman-Maldacena 
surface entropy and $S_{island}$ denotes the island entropy 
\cite{Hartman:2013qma}.   In the previous section we showed that gravitational calculations and the field 
theory of S$_{island}=$S$_{bdy}$ and S$_{HM}=$S$_{bulk}$ coincide. Hence, we 
can now describe the entanglement entropy given by the minimum in 
$S=min(S_{HM},S_{island})$. Using our analytical results  we proceed to a 
similar analysis on the nature of time and Page angle in our AdS$_3$ 
configuration \cite{Basu:2023jtf,Geng:2022dua}.

To find the Page time, we need 
to impose that $S_{HM}(t_{Page})=S_{island}$, which provides
\begin{eqnarray}
t_{Page} =\frac{\beta }{4\pi } cosh^{-1}\left[\frac{e^{-2\rho _{*}} r_{L} r_{R} 
-r_{h}( \Delta _{L} +\Delta _{R})}{2\omega ( \rho _{*}) r_{h}\sqrt{\Delta 
_{L}\Delta _{R}}}\right],
\end{eqnarray}
and considering the special case $r_{L}=r_{R}=r$  we have 
\begin{eqnarray}
&&t_{Page} =\frac{\beta }{4\pi } cosh^{-1}\left[\frac{e^{-2\rho _{*}} 
r^{2}}{2\omega ( \rho _{*}) r_{h} \Delta _{r}} -\frac{1}{\omega ( \rho 
_{*})}\right],\\
&&\Delta _{r} =r_{h} -r;\,\omega ( \rho _{*}) =1+\left(\frac{\alpha -3\gamma 
}{\gamma }\right)\cosh^{2}( \rho _{*}).
\end{eqnarray}
\begin{figure}[!]
    \centering
    \includegraphics[scale=0.6]{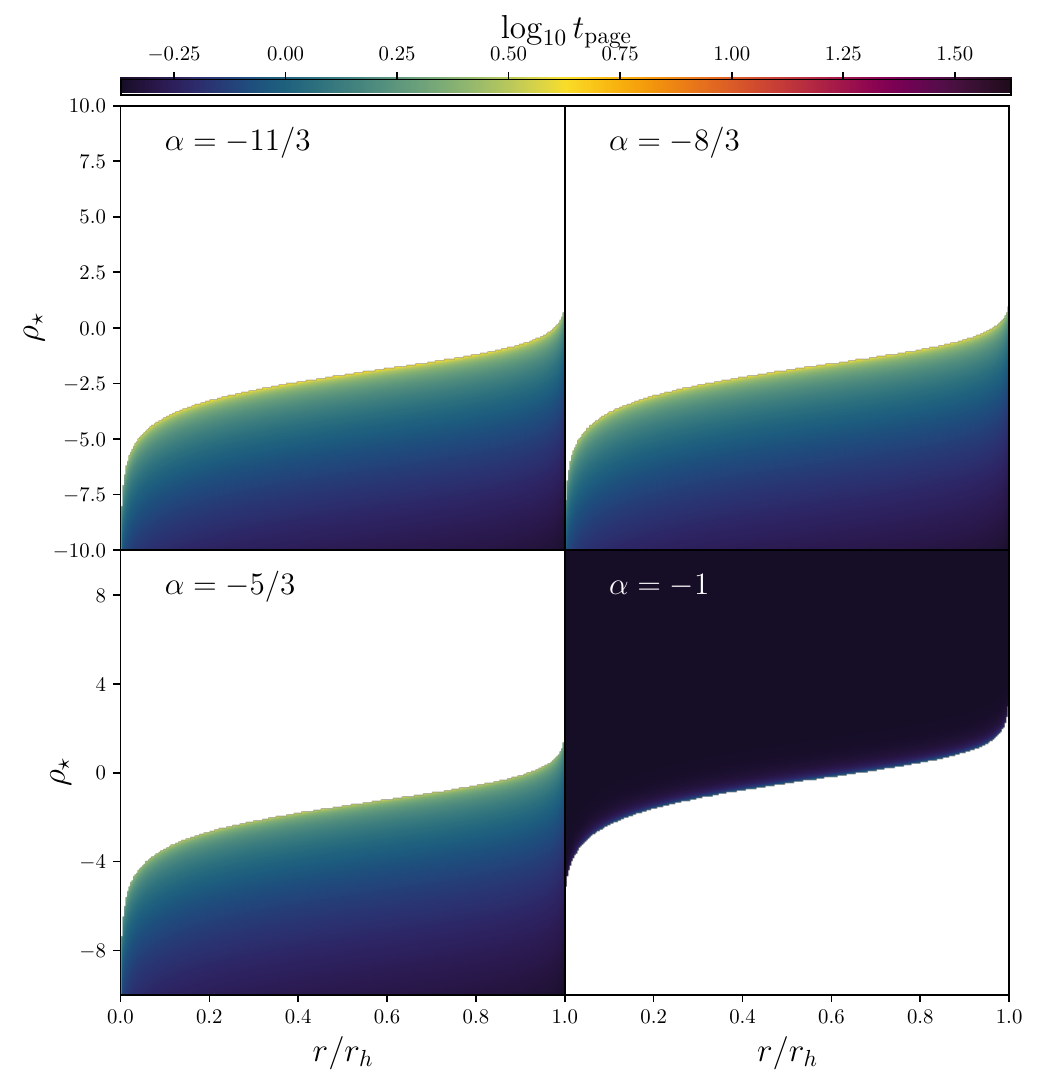}
    \caption{{\it{The Page angle    $\rho_{Page}$ for various values of 
$\alpha$ and fixed $\gamma=-0.4$. The points where the Page angle gives the tensionless brane $\rho=0$ are those where the parameter 
$\alpha$ is furthest from zero. The fact that $\alpha$  is far 
from its null value shows that for $\rho_{Page}>0$ the island dominates at $t=0$, 
while for $\rho_{Page}<0$ the HM surface dominates at $t=0$. The
regions showing dominant competition between the island and HM
surfaces initially dominate the entropy calculation due to     fixing 
$\gamma$ and varying $\alpha$, and reveal the strength of the scalar field contribution. 
These graphs are equivalent  with the density plot along a diagonal slice in 
Fig. \ref{fig:rho_page_alpha} below, with $r_L=r_R$.}}}
    \label{fig:t_page_alpha}
\end{figure}

Now, in order to find the Page angle, we consider the condition 
$S_{HM}(t=0)=S_{island}$, 
which yields
\begin{eqnarray}
\rho _{Page} =\frac{1}{2} log\left[\frac{r_{L} r_{R}}{r_{h}\left(\Delta _{L} 
+\Delta _{R} +2\left[ 1+\left(\frac{\alpha -3\gamma }{\gamma 
}\right)\right]\sqrt{\Delta _{L} \Delta _{R}}\right)}\right],
\end{eqnarray}
and for the special case $r_{L}=r_{R}=r$  we result to the  expression
\begin{eqnarray}
\rho _{Page} =\frac{1}{2} log\left\{\frac{r^{2}}{r_{h}\left( 2\Delta _{r} 
+2\left[ 1+\left(\frac{\alpha -3\gamma }{\gamma }\right)\right] \Delta 
_{r}\right)}\right\}.
\end{eqnarray}


\begin{figure}[!]
    \centering
    \includegraphics[scale=0.6]{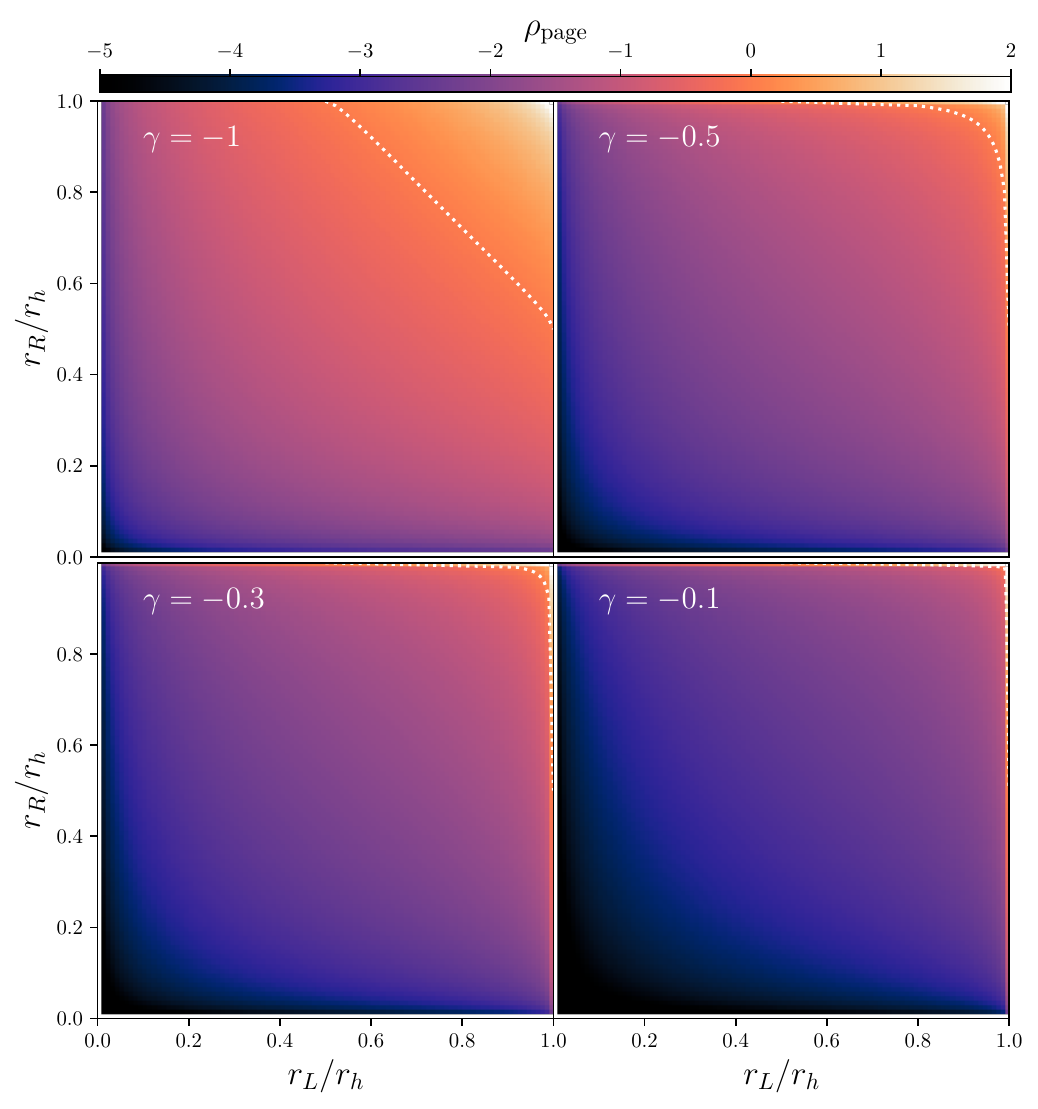}
    \caption{{\it{The  evolution of the density of the Page 
angle 
$\rho_{Page}$ around $r_L$ and $r_R$, for various values of 
$\gamma$ and fixed $\alpha=-8/3$.
The contour with $\rho_{Page}=0$ is depicted by a white dotted curve as a 
reference.}}}
    \label{fig:rho_page_gamma}
\end{figure}

In Figs. \ref{fig:t_page_gamma} and \ref{fig:t_page_alpha} we  depict the 
evolution of the Page angle density $\rho_{Page}$ as a 
function of $r_L/r_h$ and $r_R/r_h$, where $r_L$ and $r_R$ are     the positions 
of the left and right boundaries, for various values of 
$\gamma$ and   $\alpha$. These figures demonstrate that the 
Horndeski parameters can induce significant changes in the system, particularly 
when the radial coordinate $r$ is near $0$ or the horizon radius $r_h$ 
\cite{Geng:2022dua}. Such changes align with the understanding that Horndeski 
gravity causes a   shift in the islands connected by the 
Hartman-Maldacena surface \cite{Hartman:2013qma}. Our findings reveal that the residual 
information, denoted as $S^{L}_{bdry}$ and $S^{R}_{bdry}$, significantly 
affects the location of the bipartition, except when it is near the event 
horizon or a defect. Moreover, these figures   illustrate the 
dependence of the Page curve behavior on the Horndeski parameters $\alpha$ and 
$\gamma$, as well as on the geometric factors $r_L$ and $r_R$.  
Finally, they provide visual evidence of the significant role that Horndeski gravity plays in modifying the 
dynamics of entanglement entropy compared to standard general relativity.

\begin{figure}[!htbp]
    \centering
  \includegraphics[scale=0.6]{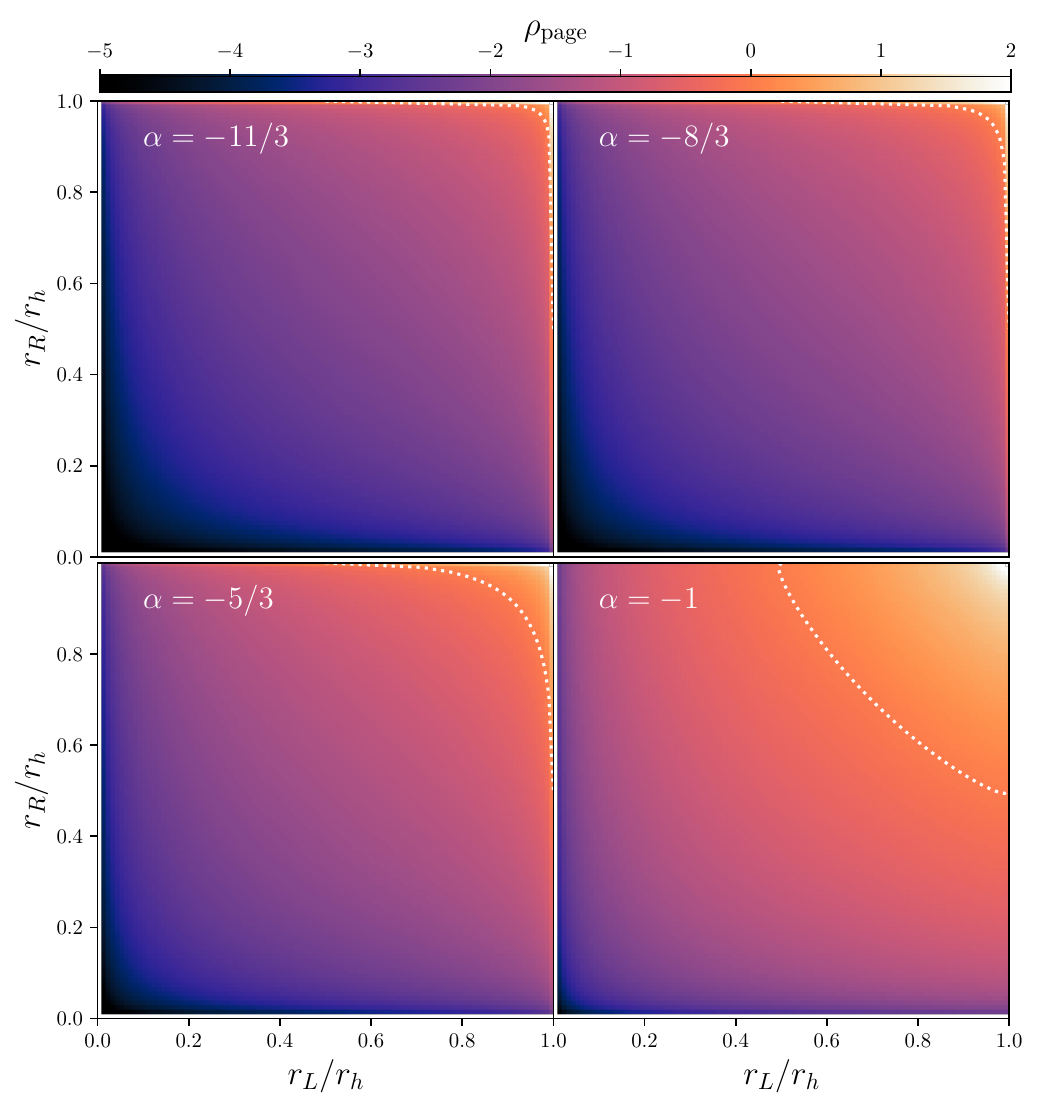}
    \caption{{\it{The       evolution of the density of the Page 
angle 
$\rho_{Page}$ about $r_L$ and $r_R$,
for various values of 
$\alpha$ and fixed $\gamma=-0.4$.  
The contour with $\rho_{Page}=0$ is depicted by a white dotted curve as a 
reference.}}}
    \label{fig:rho_page_alpha}
\end{figure}

In  Figs. \ref{fig:rho_page_gamma} and \ref{fig:rho_page_alpha} we present the  
evolution of the density of the Page angle 
$\rho_{Page}$ about $r_L$ and $r_R$, for various values of 
$\gamma$ and   $\alpha$. The horizontal axis represents $r/r_h$, ranging 
from 0 to 1,   the vertical
axis shows the Page angle $\rho_{\star}$, and the color gradient 
indicates the logarithm of the Page time. The white dotted curves depict 
the contour where $\rho_{Page}=0$, which represents the transition point between island 
dominance ($\rho_{Page}>0$) and HM surface dominance 
($\rho_{Page}<0$) at $t = 0$. As $\gamma$ changes from $-1$ to $-0.1$, we 
observe significant changes in the density patterns, particularly near the 
corners of the plot where $r_L/r_h$ or $r_R/r_h$ approach $0$ or $1$. These 
changes demonstrate the strong influence of the Horndeski parameter $\gamma$ on 
the Page angle, especially in regions close to the defect or the horizon. 
  Regions where $\rho_{\star}>0$ 
indicate island dominance at $t = 0$, while $\rho_{\star} < 0$ shows 
Hartman-Maldacena surface dominance. As $\gamma$ approaches zero, we observe 
more pronounced variations in the Page angle, particularly for smaller values 
of $r/r_h$. Additionally, as $\alpha$ changes from $-11/3$ to $-1$, we 
observe significant changes in the density patterns, particularly in regions 
where $r_L/r_h$ or $r_R/r_h$ are close to $0$ or $1$. 

The above figures demonstrate that the Page time is 
highly sensitive to the bipartition point. This sensitivity can be understood as a 
consequence of the system  geometry on the  HM surface. 
Near the defect, as illustrated in Fig. \ref{ATOLL}, the lines are compressed 
by the influence of   Horndeski   scalar field, causing the metric to 
amplify the difference between the areas of the island and the HM surface. 
Conversely, when the HM surface approaches zero area (namely where both 
bipartitions simultaneously approach the event horizon) the surface extending from the event 
horizon to any boundary point retains a nonzero area. Consequently, the Page 
time experiences significant deviations when the degrees of freedom for small 
values of $\alpha$ and $\gamma$ in both bipartitions are limited, a scenario 
applicable only within the curved bottom theory.

\section{Conclusions and discussion}\label{CONCL}

In this work we investigated the interplay between 
Horndeski gravity, boundary conditions, and entanglement entropy, particularly within the 
Anti-de-Sitter/Boundary Conformal Field Theory (AdS/BCFT) 
correspondence framework. The modifications to the Page curve due to 
Horndeski gravity reveal the   role of geometry in black hole 
information dynamics, potentially shedding light on the black hole information 
paradox. Moreover, the concept of entanglement islands, especially in the 
context of BCFT, can be helpful in examining quantum information 
distribution in these systems 
\cite{Andrzejewski:2023dja,MohammadiMozaffar:2024uyp}.  As we discussed, 
the dependence of the Page time on Horndeski parameters can serve as a 
distinguishing feature for different gravity theories, opening the way for 
possible observational signatures.

Our analysis revealed that the Page curve depends on the Horndeski 
parameters $\alpha$ and $\gamma$.  The fact that the Page time exhibits 
significant deviations when the number of degrees of freedom is small indicates 
the sensitivity of the entanglement structure to these parameters, particularly in curved backgrounds. 
This suggests that gravitational modifications can alter the fundamental 
aspects of quantum information dynamics. Our comparison with the  case where 
the boundary is flat, where the HM surface area remains constant for symmetric 
bipartitions, reveals how the geometry affects the entanglement landscape, and in particular we saw that in 
curved geometries the interplay between the horizon and the boundary can lead to richer and more 
complex behaviors.  Hence, investigating these dynamics in 
various geometries and parameter regimes could further elucidate the 
connections between gravity, quantum information, and holography. 

In summary, we showed that  Horndeski parameters significantly alter the 
behavior of the Page curve compared to standard general relativity, a feature 
caused  by the nontrivial geometry induced by the scalar Horndeski field.  
Interestingly enough,  the geometry far from the AdS limit  plays a more 
significant role comparing to  previous studies. This suggests that Horndeski gravity 
introduces important modifications to the distribution of quantum information in the 
holographic model. Lastly, we mention that   the holographic consistency  can be
  used reversely to impose  constraints on Horndeski gravity itself, providing a   new tool 
for probing the validity of modified gravity theories, establishing a novel 
connection between holography and the structure of viable gravitational theories.

\acknowledgments
The authos would like to  thank Hao Geng for the fruitful discussions. ENS 
acknowledges the contribution of     the LISA CosWG, and of COST Actions
  CA21136 ``Addressing
observational tensions in cosmology with systematics and
fundamental physics (CosmoVerse)'' and CA23130 ``Bridging high
and low energies in search of quantum gravity (BridgeQG)''.


\begin{thebibliography}{99}

\bibitem{Hawking:1975vcx}
S.~W.~Hawking,
{\it Particle Creation by Black Holes},'
Commun. Math. Phys. \textbf{43}, 199-220 (1975)
[erratum: Commun. Math. Phys. \textbf{46}, 206 (1976)].

\bibitem{Hawking:1976ra}
S.~W.~Hawking,
{\it Breakdown of Predictability in Gravitational Collapse},
Phys. Rev. D \textbf{14}, 2460-2473 (1976).
\bibitem{Susskind:2014rva}
L.~Susskind,
{\it {Computational Complexity and Black Hole Horizons,}}
Fortsch. Phys. \textbf{64}, 24-43 (2016)
[arXiv:1402.5674 [hep-th]]; (Addendum) {\sl ibid.} 44-48 
[arXiv:1403.5695 [hep-th]].

\bibitem{Brown:2015bva}
A.~R.~Brown, D.~A.~Roberts, L.~Susskind, B.~Swingle and Y.~Zhao,
{\it Holographic Complexity Equals Bulk Action?},
Phys. Rev. Lett. \textbf{116}, no.19, 191301 (2016)
[arXiv:1509.07876 [hep-th]].

\bibitem{Lloyd:2000cry}
S.~Lloyd,
{\it Ultimate physical limits to computation},
Nature \textbf{406}, 1047-1054 (2000)

\bibitem{Brown:2015lvg}
A.~R.~Brown, D.~A.~Roberts, L.~Susskind, B.~Swingle and Y.~Zhao,
{\it Complexity, action, and black holes},
Phys. Rev. D \textbf{93}, no.8, 086006 (2016)
[arXiv:1512.04993 [hep-th]].

\bibitem{Susskind:2018fmx}
L.~Susskind,
{\it Black Holes and Complexity Classes},
[arXiv:1802.02175 [hep-th]].

\bibitem{Brown:2018bms}
A.~R.~Brown, H.~Gharibyan, H.~W.~Lin, L.~Susskind, L.~Thorlacius and Y.~Zhao,
{\it Complexity of Jackiw-Teitelboim gravity},
Phys. Rev. D \textbf{99}, no.4, 046016 (2019)
[arXiv:1810.08741 [hep-th]].


\bibitem{Brown:2017jil}
A.~R.~Brown and L.~Susskind,
{\it Second law of quantum complexity},
Phys. Rev. D \textbf{97}, no.8, 086015 (2018)
[arXiv:1701.01107 [hep-th]].

\bibitem{Brown:2019whu}
A.~R.~Brown and L.~Susskind,
{\it Complexity geometry of a single qubit},
Phys. Rev. D \textbf{100}, no.4, 046020 (2019)
[arXiv:1903.12621 [hep-th]].

\bibitem{Brown:2022rwi}
A.~R.~Brown and L.~Susskind,
{\it holographic wormhole traversed in a quantum computer},
Nature \textbf{612}, no.7938, 41-42 (2022)

\bibitem{Horndeski:1974wa}
G.~W.~Horndeski,
{\it Second-order scalar-tensor field equations in a four-dimensional space,}
Int. J. Theor. Phys. \textbf{10}, 363-384 (1974).

\bibitem{Kobayashi:2019hrl}
T.~Kobayashi,
{\it Horndeski theory and beyond: a review,}
Rept. Prog. Phys. \textbf{82}, no.8, 086901 (2019)
[arXiv:1901.07183 [gr-qc]].


\bibitem{Lu:2020iav}
H.~Lu and Y.~Pang,
{\it Horndeski gravity as $D \rightarrow 4$ limit of Gauss-Bonnet,}
Phys. Lett. B \textbf{809}, 135717 (2020)
[arXiv:2003.11552 [gr-qc]].

\bibitem{Kase:2018aps}
R.~Kase and S.~Tsujikawa,
{\it Dark energy in Horndeski theories after GW170817: A review,}
Int. J. Mod. Phys. D \textbf{28}, no.05, 1942005 (2019)
[arXiv:1809.08735 [gr-qc]].

\bibitem{Koyama:2013paa}
K.~Koyama, G.~Niz and G.~Tasinato,
{\it Effective theory for the Vainshtein mechanism from the Horndeski action,}
Phys. Rev. D \textbf{88}, 021502 (2013)
[arXiv:1305.0279 [hep-th]].


\bibitem{Bahamonde:2021dqn}
S.~Bahamonde, M.~Caruana, K.~F.~Dialektopoulos, V.~Gakis, M.~Hohmann, J.~Levi 
Said, E.~N.~Saridakis and J.~Sultana,
{\it Gravitational-wave propagation and polarizations in the teleparallel 
analog of Horndeski gravity,}
Phys. Rev. D \textbf{104}, no.8, 084082 (2021)
[arXiv:2105.13243 [gr-qc]].


\bibitem{Petronikolou:2021shp}
M.~Petronikolou, S.~Basilakos and E.~N.~Saridakis,
{\it  Alleviating H0 tension in Horndeski gravity,}
[arXiv:2110.01338 [gr-qc]].

\bibitem{CANTATA:2021asi}
E.~N.~Saridakis \textit{et al.} [CANTATA],
{\it Modified Gravity and Cosmology. An Update by the CANTATA Network,}
Springer, (2021),
ISBN 978-3-030-83714-3, 978-3-030-83717-4, 978-3-030-83715-0.
[arXiv:2105.12582 [gr-qc]].


\bibitem{Abdalla:2022yfr}
E.~Abdalla, G.~Franco Abell\'an, A.~Aboubrahim, A.~Agnello, O.~Akarsu, 
Y.~Akrami, G.~Alestas, D.~Aloni, L.~Amendola and L.~A.~Anchordoqui, \textit{et 
al.}
{\it Cosmology intertwined: A review of the particle physics, astrophysics, and 
cosmology associated with the cosmological tensions and anomalies,}
JHEAp \textbf{34}, 49-211 (2022)
[arXiv:2203.06142 [astro-ph.CO]].

\bibitem{Capozziello:2011et}
S.~Capozziello and M.~De Laurentis,
{\it Extended Theories of Gravity,}
Phys. Rept. \textbf{509}, 167-321 (2011)
[arXiv:1108.6266 [gr-qc]].

\bibitem{Cai:2015emx}
Y.~F.~Cai, S.~Capozziello, M.~De Laurentis and E.~N.~Saridakis,
{\it f(T) teleparallel gravity and cosmology,}
Rept. Prog. Phys. \textbf{79}, no.10, 106901 (2016)
[arXiv:1511.07586 [gr-qc]].



\bibitem{Santos:2021orr}
F.~F.~Santos, E.~F.~Capossoli and H.~Boschi-Filho,
{\it AdS/BCFT correspondence and BTZ black hole thermodynamics within Horndeski 
gravity},
Phys. Rev. D \textbf{104}, no.6, 066014 (2021)
[arXiv:2105.03802 [hep-th]].

\bibitem{Sokoliuk:2022llp}
O.~Sokoliuk, F.~F.~Santos and A.~Baransky,
{\it AdS/BCFT correspondence and Lovelock theory in the presence of canonical 
scalar field},
[arXiv:2206.04054 [hep-th]].


\bibitem{Santos:2023flb}
F.~F.~Santos, M.~Bravo-Gaete, O.~Sokoliuk and A.~Baransky,
{\it AdS/BCFT correspondence and Horndeski gravity in the presence of gauge 
fields: holographic paramagnetism/ferromagnetism phase transition},
[arXiv:2301.03121 [hep-th]].


\bibitem{Santos:2023mee}
F.~F.~Santos, M.~Bravo-Gaete, M.~M.~Ferreira and R.~Casana,
{\it Magnetized AdS/BCFT Correspondence in Horndeski Gravity},
Fortsch. Phys. \textbf{72}, no.7-8, 2400088 (2024)
doi:10.1002/prop.202400088
[arXiv:2310.17092 [hep-th]].


\bibitem{Santos:2024zoh}
F.~F.~Santos and H.~Boschi-Filho,
{\it Holographic complexity and residual entropy of a rotating charged BTZ 
black 
hole within Horndeski gravity},
[arXiv:2407.10004 [hep-th]].


  \bibitem{Maldacena:1997re} 
  J.~M.~Maldacena,
  {\it The Large N limit of superconformal field theories and supergravity},
  Int.\ J.\ Theor.\ Phys.\  {\bf 38}, 1113 (1999)
  [Adv.\ Theor.\ Math.\ Phys.\  {\bf 2}, 231 (1998)]
    [hep-th/9711200].


\bibitem{Witten:1998qj}
E.~Witten,
{\it { Anti-de Sitter space and holography,}}
Adv. Theor. Math. Phys. \textbf{2} (1998), 253-291
[arXiv:hep-th/9802150 [hep-th]].

\bibitem{Caceres:2023gfa}
N.~Caceres, C.~Corral, F.~Diaz and R.~Olea,
{\it Holographic renormalization of Horndeski gravity},
JHEP \textbf{05}, 125 (2024)
doi:10.1007/JHEP05(2024)125
[arXiv:2311.04054 [hep-th]].




\bibitem{Santos:2024cwf}
F.~F.~Santos and H.~Boschi-Filho,
{\it Geometric Josephson junction},
[arXiv:2407.10008 [hep-th]].

\bibitem{Takayanagi:2011zk} 
T.~Takayanagi,
{\it { Holographic Dual of BCFT,}}
Phys.\ Rev.\ Lett.\  {\bf 107}, 101602 (2011),
[arXiv:1105.5165 [hep-th]].

\bibitem{Fujita:2011fp}{ 
M.~Fujita, T.~Takayanagi and E.~Tonni,
{\it Aspects of AdS/BCFT},
JHEP {\bf 1111}, 043 (2011),
[arXiv:1108.5152 [hep-th]].}

\bibitem{Kanda:2023zse}
H.~Kanda, M.~Sato, Y.~k.~Suzuki, T.~Takayanagi and Z.~Wei,
{\it AdS/BCFT with brane-localized scalar field},
JHEP \textbf{03}, 105 (2023)
[arXiv:2302.03895 [hep-th]].

\bibitem{dosSantos:2022scy}
F.~F.~dos Santos,
{\it AdS/BCFT correspondence and BTZ black hole within electric field},
JHAP \textbf{4}, no.1, 81-92 (2022)
[arXiv:2206.09502 [hep-th]].



\bibitem{Ryu:2006bv}
S.~Ryu and T.~Takayanagi,
{\it Holographic derivation of entanglement entropy from AdS/CFT},
Phys. Rev. Lett. \textbf{96}, 181602 (2006)
[arXiv:hep-th/0603001 [hep-th]].



\bibitem{DosSantos:2022exb}
F.~F.~Dos Santos,
{\it Entanglement entropy in Horndeski gravity},
JHAP \textbf{3}, no.1, 1-14 (2022)
[arXiv:2201.02500 [hep-th]].

\bibitem{Caceres:2017lbr}
E.~Caceres, R.~Mohan and P.~H.~Nguyen,
{\it On holographic entanglement entropy of Horndeski black holes},
JHEP \textbf{10}, 145 (2017)
doi:10.1007/JHEP10(2017)145
[arXiv:1707.06322 [hep-th]].

\bibitem{Feng:2015oea}
X.~H.~Feng, H.~S.~Liu, H.~L\"u and C.~N.~Pope,
{\it Black Hole Entropy and Viscosity Bound in Horndeski Gravity},
JHEP \textbf{11}, 176 (2015)
doi:10.1007/JHEP11(2015)176
[arXiv:1509.07142 [hep-th]].

\bibitem{Wald:1993nt}
R.~M.~Wald,
{\it Black hole entropy is the Noether charge},
Phys. Rev. D \textbf{48}, no.8, R3427-R3431 (1993)
doi:10.1103/PhysRevD.48.R3427
[arXiv:gr-qc/9307038 [gr-qc]].

\bibitem{Iyer:1994ys}
V.~Iyer and R.~M.~Wald,
{\it Some properties of Noether charge and a proposal for dynamical black hole entropy},
Phys. Rev. D \textbf{50}, 846-864 (1994)
doi:10.1103/PhysRevD.50.846
[arXiv:gr-qc/9403028 [gr-qc]].


\bibitem{Karch:2000ct}
A.~Karch and L.~Randall,
{\it Locally localized gravity},
JHEP \textbf{05}, 008 (2001)
[arXiv:hep-th/0011156 [hep-th]].

\bibitem{DeWolfe:2001pq}
O.~DeWolfe, D.~Z.~Freedman and H.~Ooguri,
{\it { Holography and defect conformal field theories,}}
Phys. Rev. D \textbf{66}, 025009 (2002)
[arXiv:hep-th/0111135 [hep-th]].

\bibitem{Bak:2003jk}
D.~Bak, M.~Gutperle and S.~Hirano,
{\it{ A Dilatonic deformation of AdS(5) and its field theory dual,}}
JHEP \textbf{05}, 072 (2003)
[arXiv:hep-th/0304129 [hep-th]].

\bibitem{Clark:2004sb}
A.~B.~Clark, D.~Z.~Freedman, A.~Karch and M.~Schnabl,
{\it {Dual of the Janus solution: An interface conformal field theory,}}
Phys. Rev. D \textbf{71}, 066003 (2005)
[arXiv:hep-th/0407073 [hep-th]].


\bibitem{Cardy:2004hm}
J.~L.~Cardy,
{\it Boundary conformal field theory,}
[arXiv:hep-th/0411189 [hep-th]].

\bibitem{Tonni:2010pv}
E.~Tonni,
{\it Holographic entanglement entropy: near horizon geometry and disconnected regions},
JHEP \textbf{05} (2011), 004,
[arXiv:1011.0166 [hep-th]].


\bibitem{Azeyanagi:2007qj}
T.~Azeyanagi, A.~Karch, T.~Takayanagi and E.~G.~Thompson,
{\it { Holographic calculation of boundary entropy,}}
JHEP \textbf{03}, 054 (2008)
[arXiv:0712.1850 [hep-th]].

\bibitem{Setare:2008hm}
M.~R.~Setare and E.~N.~Saridakis,
{\it{ Correspondence between Holographic and Gauss-Bonnet dark energy models,}}
Phys. Lett. B \textbf{670}, 1-4 (2008),
[arXiv:0810.3296 [hep-th]].

\bibitem{Saridakis:2017rdo}
E.~N.~Saridakis,
{\it { Ricci-Gauss-Bonnet holographic dark energy,}}
Phys. Rev. D \textbf{97}, no.6, 064035 (2018),
[arXiv:1707.09331 [gr-qc]].


 
\bibitem{Basilakos:2023seo}
S.~Basilakos, A.~Lymperis, M.~Petronikolou and E.~N.~Saridakis,
{\it Barrow holographic dark energy with varying exponent,}
[arXiv:2312.15767 [gr-qc]].



\bibitem{Almheiri:2019hni}
A.~Almheiri, R.~Mahajan, J.~Maldacena and Y.~Zhao,
{\it The Page curve of Hawking radiation from semiclassical geometry},
JHEP \textbf{03}, 149 (2020)
[arXiv:1908.10996 [hep-th]].

\bibitem{Rozali:2019day}
M.~Rozali, J.~Sully, M.~Van Raamsdonk, C.~Waddell and D.~Wakeham,
{\it Information radiation in BCFT models of black holes},
JHEP \textbf{05}, 004 (2020)
[arXiv:1910.12836 [hep-th]].

\bibitem{Chen:2020uac}
H.~Z.~Chen, R.~C.~Myers, D.~Neuenfeld, I.~A.~Reyes and J.~Sandor,
{\it Quantum Extremal Islands Made Easy, Part I: Entanglement on the Brane},
JHEP \textbf{10}, 166 (2020)
[arXiv:2006.04851 [hep-th]].

\bibitem{Chen:2020hmv}
H.~Z.~Chen, R.~C.~Myers, D.~Neuenfeld, I.~A.~Reyes and J.~Sandor,
{\it Quantum Extremal Islands Made Easy, Part II: Black Holes on the Brane},
JHEP \textbf{12}, 025 (2020)
[arXiv:2010.00018 [hep-th]].

\bibitem{Deng:2020ent}
F.~Deng, J.~Chu and Y.~Zhou,
{\it Defect extremal surface as the holographic counterpart of Island formula},
JHEP \textbf{03}, 008 (2021)
[arXiv:2012.07612 [hep-th]].

\bibitem{Suzuki:2022xwv}
K.~Suzuki and T.~Takayanagi,
{\it BCFT and Islands in two dimensions},
JHEP \textbf{06}, 095 (2022)
[arXiv:2202.08462 [hep-th]].

\bibitem{Geng:2021iyq}
H.~Geng, S.~L\"ust, R.~K.~Mishra and D.~Wakeham,
{\it Holographic BCFTs and Communicating Black Holes},
jhep \textbf{08}, 003 (2021)
[arXiv:2104.07039 [hep-th]].

\bibitem{Geng:2024xpj}
H.~Geng,
{\it Replica Wormholes and Entanglement Islands in the Karch-Randall Braneworld},
[arXiv:2405.14872 [hep-th]].


\bibitem{Geng:2020fxl}
H.~Geng, A.~Karch, C.~Perez-Pardavila, S.~Raju, L.~Randall, M.~Riojas and 
S.~Shashi,
{\it Information Transfer with a Gravitating Bath},
SciPost Phys. \textbf{10}, no.5, 103 (2021)
[arXiv:2012.04671 [hep-th]].

\bibitem{Ryu:2006ef}
S.~Ryu and T.~Takayanagi,
{\it Aspects of Holographic Entanglement Entropy},
JHEP \textbf{08} (2006), 045,
[arXiv:hep-th/0605073 [hep-th]].

\bibitem{Basu:2023jtf}
D.~Basu, H.~Chourasiya, V.~Raj and G.~Sengupta,
{\it Reflected entropy in a BCFT on a black hole background},
JHEP \textbf{05}, 054 (2024)
[arXiv:2311.17023 [hep-th]].


\bibitem{Randall:1999vf}
L.~Randall and R.~Sundrum,
{\it An Alternative to compactification},
Phys. Rev. Lett. \textbf{83}, 4690-4693 (1999)
[arXiv:hep-th/9906064 [hep-th]].

\bibitem{Brito:2018pwe}
F.~A.~Brito and F.~F.~Santos,
{\it Braneworlds in Horndeski gravity},
Eur. Phys. J. Plus \textbf{137}, no.9, 1051 (2022)
[arXiv:1810.08196 [hep-th]].

\bibitem{Santos:2023eqp}
F.~F.~Santos, B.~Pourhassan and E.~Saridakis,
{\it de Sitter versus anti-de Sitter in Horndeski-like gravity},
Fortsch. Phys. \textbf{72}, no.3, 2300228 (2024)
[arXiv:2305.05794 [hep-th]].

\bibitem{Almheiri:2019psf}
A.~Almheiri, N.~Engelhardt, D.~Marolf and H.~Maxfield,
{\it The entropy of bulk quantum fields and the entanglement wedge of an 
evaporating black hole},
JHEP \textbf{12}, 063 (2019)
[arXiv:1905.08762 [hep-th]].

\bibitem{Penington:2019npb}
G.~Penington,
{\it Entanglement Wedge Reconstruction and the Information Paradox},
JHEP \textbf{09}, 002 (2020)
[arXiv:1905.08255 [hep-th]].


\bibitem{Engelhardt:2014gca}
N.~Engelhardt and A.~C.~Wall,
{\it Quantum Extremal Surfaces: Holographic Entanglement Entropy beyond the 
Classical Regime},
JHEP \textbf{01}, 073 (2015)
[arXiv:1408.3203 [hep-th]].

\bibitem{Almheiri:2019psy}
A.~Almheiri, R.~Mahajan and J.~E.~Santos,
{\it Entanglement islands in higher dimensions},
SciPost Phys. \textbf{9}, no.1, 001 (2020)
[arXiv:1911.09666 [hep-th]].

\bibitem{Hartman:2013qma}
T.~Hartman and J.~Maldacena,
{\it Time Evolution of Entanglement Entropy from Black Hole Interiors},
JHEP \textbf{05}, 014 (2013)
[arXiv:1303.1080 [hep-th]].

\bibitem{Geng:2021mic}
H.~Geng, A.~Karch, C.~Perez-Pardavila, S.~Raju, L.~Randall, M.~Riojas and 
S.~Shashi,
{\it Entanglement phase structure of a holographic BCFT in a black hole 
background},
JHEP \textbf{05}, 153 (2022)
[arXiv:2112.09132 [hep-th]].


\bibitem{Geng:2023qwm}
H.~Geng,
{\it Revisiting Recent Progress in the Karch-Randall Braneworld},
[arXiv:2306.15671 [hep-th]].

\bibitem{Geng:2021hlu}
H.~Geng, A.~Karch, C.~Perez-Pardavila, S.~Raju, L.~Randall, M.~Riojas and 
S.~Shashi,
{\it Inconsistency of islands in theories with long-range gravity},
JHEP \textbf{01}, 182 (2022)
[arXiv:2107.03390 [hep-th]].

\bibitem{Geng:2023zhq}
H.~Geng,
{\it Graviton Mass and Entanglement Islands in Low Spacetime Dimensions},
[arXiv:2312.13336 [hep-th]].

\bibitem{Geng:2022dua}
H.~Geng, L.~Randall and E.~Swanson,
{\it BCFT in a black hole background: an analytical holographic model},
JHEP \textbf{12}, 056 (2022)
[arXiv:2209.02074 [hep-th]].

\bibitem{Bass:2021acr}
S.~D.~Bass, A.~De Roeck and M.~Kado,
{\it The Higgs boson implications and prospects for future discoveries},
Nature Rev. Phys. \textbf{3}, no.9, 608-624 (2021)
[arXiv:2104.06821 [hep-ph]].

\bibitem{Higgs2014}
Higgs, P. W. {\it Nobel Lecture: Evading the Goldstone theorem}.
Rev. Mod. Phys. 86, 851, 

\bibitem{Jeong:2022zea}
H.~S.~Jeong, K.~Y.~Kim and Y.~W.~Sun,
{\it Holographic entanglement density for spontaneous symmetry breaking},
JHEP \textbf{06}, 078 (2022)
[arXiv:2203.07612 [hep-th]].

\bibitem{Saridakis:2007ns}
E.~N.~Saridakis,
{\it Holographic Dark Energy in Braneworld Models with Moving Branes and the 
w=-1 Crossing,}
JCAP \textbf{04}, 020 (2008)
[arXiv:0712.2672 [astro-ph]].


\bibitem{Kofinas:2014qxa}
G.~Kofinas, E.~N.~Saridakis and J.~Q.~Xia,
{\it Cosmological solutions and observational constraints on five-dimensional 
braneworld cosmology with gravitating Nambu-Goto matching conditions,}
Phys. Rev. D \textbf{90}, no.8, 084049 (2014)
[arXiv:1403.7510 [astro-ph.CO]].
\bibitem{Li:2018kqp}
Y.~Z.~Li and H.~Lu,
{\it $a$-theorem for Horndeski gravity at the critical point},
Phys. Rev. D \textbf{97}, no.12, 126008 (2018)
[arXiv:1803.08088 [hep-th]].


\bibitem{Dvali:2000hr}
G.~R.~Dvali, G.~Gabadadze and M.~Porrati,
{\it 4-D gravity on a brane in 5-D Minkowski space},
Phys. Lett. B \textbf{485}, 208-214 (2000)
[arXiv:hep-th/0005016 [hep-th]].

\bibitem{Heisenberg:2018vsk}
L.~Heisenberg,
{\it A systematic approach to generalisations of General Relativity and their 
cosmological implications},
Phys. Rept. \textbf{796}, 1-113 (2019)
[arXiv:1807.01725 [gr-qc]].


\bibitem{Doroudiani:2019llj}
M.~Doroudiani, A.~Naseh and R.~Pirmoradian,
{\it Complexity for Charged Thermofield Double States},
JHEP \textbf{01}, 120 (2020)
[arXiv:1910.08806 [hep-th]].

\bibitem{Iliesiu:2022kny}
L.~V.~Iliesiu, S.~Murthy and G.~J.~Turiaci,
{\it Black hole microstate counting from the gravitational path integral},
[arXiv:2209.13602 [hep-th]].

\bibitem{Balasubramanian:2022gmo}
V.~Balasubramanian, A.~Lawrence, J.~M.~Magan and M.~Sasieta,
{\it Microscopic Origin of the Entropy of Black Holes in General Relativity},
Phys. Rev. X \textbf{14}, no.1, 011024 (2024)
[arXiv:2212.02447 [hep-th]].

\bibitem{Andrzejewski:2023dja}
K.~Andrzejewski,
{\it Evolution of capacity of entanglement and modular entropy in harmonic 
chains and scalar fields},
Phys. Rev. D \textbf{108}, no.12, 125013 (2023)
[arXiv:2309.03013 [hep-th]].

\bibitem{MohammadiMozaffar:2024uyp}
M.~R.~Mohammadi Mozaffar,
{\it Capacity of entanglement for scalar fields in squeezed states},
Phys. Rev. D \textbf{110}, no.4, 046021 (2024)
[arXiv:2405.09128 [hep-th]].

 
 



\end{thebibliography}
\end{document}